%% file: ANA-TOPQ-2023-10-PAPER.tex
\pdfoutput=1
\documentclass[PAPER, atlasdraft=false, cernpreprint, UKenglish, texmf, orcidlogo]{atlasdoc}
\usepackage{atlaspackage}
\usepackage{atlasbiblatex}

\usepackage{atlasphysics}
\graphicspath{{logos/}{figures/}}

\usepackage{ANA-TOPQ-2023-10-PAPER-defs}
\usepackage{dcolumn}
\usepackage{atlas_doc-defs}
\usepackage{atlas_tables-defs}
\usepackage{bm}

\AtlasTitle{Test of lepton flavour universality in $W$-boson decays into electrons and $\tau$-leptons using $pp$ collisions at $\sqrt{s}=13$ TeV with the ATLAS detector}

\AtlasVersion{2.0}

\AtlasAbstract{%
A measurement of the ratio of the branching fractions, $R_{\tau/e} = B(W \to \tau \nu)/ B(W \to e \nu)$, is performed using a sample of $W$ bosons originating from top-quark decays to final states containing $\tau$-leptons or electrons.
This measurement uses $pp$ collisions at $\sqrt{s}=13$ TeV, collected by the ATLAS experiment at the Large Hadron Collider during Run 2, corresponding to an integrated luminosity of 140 fb$^{-1}$.
The $W \to \tau \nu_\tau$ (with $\tau \to e \nu_e \nu_\tau$) and $W \to e \nu_e$ decays are distinguished using the differences in the impact parameter distributions and transverse momentum spectra of the electrons. The measured ratio of branching fractions $R_{\tau/e} = 0.975 \pm 0.012~\textrm{(stat.)} \pm 0.020~\textrm{(syst.)}$, is consistent with the Standard Model assumption of lepton flavour universality in $W$-boson decays.
}

\PreprintIdNumber{CERN-EP-2024-315}

\arXivId{2412.11989}

\AtlasJournalRef{\JHEP 05 (2025) 038}
\AtlasDOI{10.1007/JHEP05(2025)038}

\hypersetup{pdftitle={ATLAS document},pdfauthor={The ATLAS Collaboration}}

\begin{document}

\maketitle

\tableofcontents

%
%

%
%

% The next lines are included from the .//sections/01_Introduction.tex input file
\section{Introduction}

The Standard Model (SM) \cite{Glashow:1961tr,Weinberg:1967tq,Salam:1968rm} posits that the
electroweak interactions of charged leptons (electrons, muons and $\tau$-leptons)
are identical.
This feature, known as lepton flavour universality (LFU), is one of the foundational principles of the SM. Differences in the rates of electroweak processes among charged leptons are induced only via their mass differences, manifesting in phase space and radiative effects. Any violation of LFU would be an unambiguous signal of physics beyond the SM and
would help to identify the principles that can be used to build a more comprehensive theory.

LFU has been extensively tested by measuring the couplings of the different generations of leptons %
to the $W$ boson. Such studies have been performed in decays of $\tau$-leptons, charged $\pi$ and $K$ mesons \cite{Pich:2013lsa}, and in leptonic decays of the $W$ bosons \cite{Workman:2022ynf}. No significant deviations from LFU have been observed in the charged-current decays of $\tau$-leptons, $\pi$ and $K$ mesons, with experimental precision reaching about 0.1--0.2\%.
The precision of LFU tests in the decays of on-shell $W$ bosons is lower. The ratios of the branching fractions of the W-boson decays into leptons are consistent with LFU, and the world-average values have a precision of 0.6--2\%.
Despite higher experimental uncertainties, probes of LFU in decays of on-shell $W$ bosons offer sensitivity to different possible new physics effects compared to the decays of light particles \cite{Crivellin:2022rhw,Allwicher:2023aql,Abada:2023raf}.

The ratio of branching fractions $\Rtaumu \equiv B(W \to \tau \nu_{\tau})/B(W \to \mu \nu_{\mu})$ was measured at the LEP collider~\cite{ALEPH:2013dgf} and at the Large Hadron Collider (LHC) by the ATLAS \cite{ATLAS:2020xea} and CMS \cite{CMS:2022mhs} collaborations.
The measurements at LEP deviate from unity by more than two standard deviations, but the more precise subsequent measurements from ATLAS and CMS agree with the SM expectation.
The Particle Data Group (PDG) average of all the measurements of \Rtaumu, accounting for the tension between them, is
$1.002 \pm 0.020$~\cite{Workman:2022ynf}.
Recently, the ATLAS Collaboration published a measurement of $R_{\mu/e} \equiv B(W \to \mu \nu_{\mu})/B(W \to e \nu_e) = 0.9995 \pm 0.0044$ \cite{ATLAS:2024tlf}
with an improved precision compared to the current world average $R_{\mu/e} = 1.002 \pm 0.006$~\cite{Workman:2022ynf}.
The combined LEP result of $\Rtaue \equiv  B(W \to \tau \nu_{\tau})/B(W \to e \nu_e) = 1.063 \pm 0.027$ deviates from the SM by more than two standard deviations like the LEP $\Rtaumu$ measurement
\cite{ALEPH:2013dgf}.
The result of the CMS Collaboration $\Rtaue  =  0.994 \pm 0.021$~\cite{CMS:2022mhs}
agrees with the SM expectation and differs from the corresponding LEP measurement.

This paper presents the first measurement of the ratio \Rtaue with the ATLAS detector using top-quark decays as a source of $W$ bosons. The analysis complements the previous ATLAS measurement of \Rtaumu and $R_{\mu/e}$ in $W$-boson decays by including another decay channel.
Top-quark pair-production yields a clean source of $W$ bosons, owing to the large \ttbar production cross-section relative to expected background processes and efficient triggering. The SM predicts that the top quark decays almost exclusively into a $Wb$ final state, and thus a measurement of the ratio of branching fractions $B(t \to \tau \nu b)/B(t \to e \nu b)$ provides a measurement of \Rtaue.
For the $\tau$-lepton selection, the leptonic decay $\tau \to e \nu_e \nu_{\tau}$ is chosen over hadronic $\tau$ decays due to the
smaller experimental uncertainties associated with the leptonic channel.
In addition, selecting the electron in both the $W\rightarrow\tau\nu_\tau$ and $W\rightarrow e\nu_e$ final states results in a substantial reduction of the systematic uncertainties in the ratio \Rtaue.

This paper is organised as follows. Section~\ref{sec:detector} presents the ATLAS detector, followed by a description of the data and Monte Carlo simulation samples used in the analysis in Section~\ref{sec:samples}. Section~\ref{sec:events} details the physical objects and event selection criteria. The calibration of the signal modelling is discussed in Section~\ref{sec:d0_cal}, and Section~\ref{sec:background} describes the estimate of background contributions. The statistical analysis and systematic uncertainties are covered in Section~\ref{sec:fit}, with the results and conclusions presented in Sections~\ref{sec:results} and \ref{sec:conclusion}, respectively.

%

% End of text imported from the .//sections/01_Introduction.tex input file

% The next lines are included from the .//sections/02_Atlas-detector.tex input file
\section{ATLAS detector}
\label{sec:detector}

\newcommand{\AtlasCoordFootnote}{%
ATLAS uses a right-handed coordinate system with its origin at the nominal interaction point (IP)
in the centre of the detector and the \(z\)-axis along the beam pipe.
The \(x\)-axis points from the IP to the centre of the LHC ring,
and the \(y\)-axis points upwards.
Polar coordinates \((r,\phi)\) are used in the transverse plane,
\(\phi\) being the azimuthal angle around the \(z\)-axis.
The pseudorapidity is defined in terms of the polar angle \(\theta\) as \(\eta = -\ln \tan(\theta/2)\) and is equal to the rapidity
$ y = \frac{1}{2} \ln \left( \frac{E + p_z}{E - p_z} \right) $ in the relativistic limit.
Angular distance is measured in units of \(\Delta R \equiv \sqrt{(\Delta y)^{2} + (\Delta\phi)^{2}}\). The transverse momentum \pT is defined as the projection of the momentum of a particle on the plane perpendicular to the beam pipe in the centre of ATLAS detector.}

The ATLAS detector~\cite{PERF-2007-01} at the LHC covers nearly the entire solid angle around the collision point.\footnote{\AtlasCoordFootnote}
It consists of an inner tracking detector surrounded by a thin superconducting solenoid, electromagnetic and hadronic calorimeters,
and a muon spectrometer incorporating three large superconducting air-core toroidal magnets.

The inner-detector system (ID) is immersed in a \qty{2}{\tesla} axial magnetic field
and provides charged-particle tracking in the range of \(|\eta| < 2.5\).
The high-granularity silicon pixel detector covers the vertex region and typically provides four measurements per track,
the first hit generally being in the insertable B-layer (IBL) installed before Run~2~\cite{ATLAS-TDR-19,PIX-2018-001}.
It is followed by the SemiConductor Tracker (SCT), which usually provides eight measurements per track.
These silicon detectors are complemented by the transition radiation tracker (TRT),
which enables radially extended track reconstruction up to \(|\eta| = 2.0\).
The TRT also provides electron identification information
based on the fraction of hits (typically 30 in total) above a higher energy-deposit threshold arising from transition radiation.

The calorimeter system covers the pseudorapidity range \(|\eta| < 4.9\).
Within the region \(|\eta|< 3.2\), electromagnetic calorimetry is provided by barrel and
endcap high-granularity lead/liquid-argon (LAr) calorimeters,
with an additional thin LAr presampler covering \(|\eta| < 1.8\)
to correct for energy loss in material upstream of the calorimeters.
Hadronic calorimetry is provided by the steel/scintillator-tile calorimeter,
segmented into three barrel structures within \(|\eta| < 1.7\), and two copper/LAr hadronic endcap calorimeters.
The solid angle coverage is completed with forward copper/LAr and tungsten/LAr calorimeter modules
optimised for electromagnetic and hadronic energy measurements, respectively.

The muon spectrometer (MS) comprises separate trigger and
high-precision tracking chambers measuring the deflection of muons in a magnetic field generated by the superconducting air-core toroidal magnets.
The field integral of the toroids ranges between \num{2.0} and \qty{6.0}{\tesla\metre}
across most of the detector.
Three layers of precision chambers, each consisting of layers of monitored drift tubes, cover the region \(|\eta| < 2.7\),
complemented by cathode-strip chambers in the forward region, where the background is highest.
The muon trigger system covers the range \(|\eta| < 2.4\) with resistive-plate chambers in the barrel, and thin-gap chambers in the endcap regions.

The luminosity is measured mainly by the LUCID--2~\cite{LUCID2} detector that records Cherenkov light produced in the quartz windows of photomultipliers located close to the beampipe.

Events are selected by the first-level trigger system implemented in custom hardware,
followed by selections made by algorithms implemented in software in the high-level trigger~\cite{TRIG-2016-01}.
The first-level trigger accepts events from the \qty{40}{\MHz} bunch crossings at a rate below \qty{100}{\kHz},
which the high-level trigger further reduces to record complete events to disk at about \qty{1}{\kHz}.

A software suite~\cite{SOFT-2022-02-update} is used in data simulation, in the reconstruction
and analysis of real and simulated data, in detector operations, and in the trigger and data acquisition
systems of the experiment.

%

%
%
%
%
%
%
%
%
%
%
%
%
%
%
%
%
%
%
%
%
%
%
%
%
%
%
%
%
%
%
%
%
%
%
%
%
%
%
%
%
%
%
%
%
%
%
%
%
%
%
%
%

% End of text imported from the .//sections/02_Atlas-detector.tex input file

% The next lines are included from the .//sections/04_Data_and_Simulation.tex input file
%
%
%
%
%
%
%
%
%
%
%
%

\section{Data and simulated events samples}
\label{sec:samples}

The measurement uses $pp$ collision data recorded by the
ATLAS detector from 2015 to 2018 at a centre-of-mass energy of $\sqrt{s} = 13$ TeV.
Each recorded event includes on average 33 additional inelastic collisions
because of the high instantaneous luminosity of the LHC. These background collisions are referred to as \textit{\pileup}. After
the application of data-quality requirements \cite{ATLAS:2019fst}, the data sample corresponds
to an integrated luminosity of $140.1 \pm 1.2$~fb$^{-1}$~\cite{ATLAS:2022hro}.

Samples of simulated events were produced using Monte Carlo (MC)
techniques to model different SM processes. After event generation
of the process of interest for each sample, the detector response
was modelled using either the full ATLAS detector simulation based on \GEANT toolkit \cite{GEANT4:2002zbu} or a fast simulation that relied on a parametrisation of the calorimeter response~\cite{ATLAS:2010arf}. The effect of \pileup was modelled by overlaying the simulated hard-scattering event with
inelastic $pp$ collisions generated with \PYTHIA[8.186]~\cite{Sjostrand:2007gs}
using the \NNPDF[2.3lo]~\cite{Ball:2012cx} set of parton distribution functions (PDF)
and the
A3 set of tuned parameters~\cite{ATL-PHYS-PUB-2016-017}.
The data and MC simulated events were passed through the same reconstruction
and analysis procedures.

For all MC samples simulating the production of top quarks,
the mass of the top quark was set to $\mtop = 172.5$ GeV. The mass of the Higgs boson was set to 125 GeV.
In all samples where \PYTHIA was used to simulate the parton shower (PS) and hadronisation, it used the A14 set of tuned parameters~\cite{ATL-PHYS-PUB-2014-021} and the \NNPDF[2.3lo] set of PDFs. In all samples where the \PYTHIA or \HERWIG~\cite{Bahr:2008pv,Bellm:2017bvx} generators were used to simulate the PS and hadronisation, the decays of bottom and charm hadrons were simulated using the \EVTGEN program~\cite{Lange:2001uf}.

The production of \ttbar events was modelled using the
\POWHEGBOX[v2]~\cite{Frixione:2007nw,Nason:2004rx,Frixione:2007vw,Alioli:2010xd}
generator, which provides matrix elements (ME) at next-to-leading
order~(NLO) in the strong coupling constant \alphas, and the
\NNPDF[3.0nlo]~\cite{Ball:2014uwa}
PDF set.
The \hdamp parameter, which
effectively regulates the high-transverse momentum (\pT) radiation against which the
\ttbar system recoils, was set to 1.5\,\mtop~\cite{ATL-PHYS-PUB-2016-020}.
The functional form of the renormalisation and factorisation scales was
set to \(\sqrt{\mtop^{2} + \pTX[2]}\).
The events were passed through
\PYTHIA[8.230]~\cite{Sjostrand:2014zea} for the PS and
hadronisation.

An additional re-weighting was applied to the \ttbar events to match the calculations at next-to-next-to-leading-order (NNLO) in \alphas
and NLO in $\alpha_{\textrm{EW}}$ \cite{Czakon_2017}.
The  applied weights are based on the parton-level distributions, after final state radiation, and are derived as a function of the top-quark and anti-top-quark \pT, and the mass and transverse momentum of the \ttbar system.
The re-weighting improves the agreement between data and MC simulation.

The \ttbar sample was normalised to the cross-section prediction at NNLO
in \alphas including the resummation of next-to-next-to-leading logarithmic (NNLL) soft-gluon terms calculated using
\TOPpp[2.0]~\cite{Beneke:2011mq,Cacciari:2011hy,Baernreuther:2012ws,Czakon:2012zr,Czakon:2012pz,Czakon:2013goa,Czakon:2011xx}.
For $pp$ collisions at a centre-of-mass energy of \(\rts = \qty{13}{\TeV}\), this cross-section corresponds to
\(\sigma(\ttbar)_{\text{NNLO+NNLL}} = 832 \pm 45\,\unit{\pb}\).
The uncertainties in the cross-section due to the PDF and \alphas were calculated using the \PDFforLHC[15] prescription~\cite{Butterworth:2015oua}
with the \MSTW[nnlo]~\cite{Martin:2009iq,Martin:2009bu}, \CT[10nnlo]~\cite{Lai:2010vv,Gao:2013xoa}
and \NNPDF[2.3lo] PDF sets in the five-flavour scheme, and were added in quadrature to the effect of the scale uncertainty.

In addition to \ttbar events, the contribution of the $Wt$ and $\ttbar+X$ $(X=W, Z, H)$ production processes are treated as signal.
For the $Wt$ process, the leptons (electron or $\tau$-lepton) coming from the top-quark and $W$-boson decays are included in the analysis as in both cases they are produced in the decays of a real $W$ boson.

Single-top \(Wt\) associated production was modelled using
\POWHEGBOX[v2] and \PYTHIA[8.230].
The
renormalisation and factorisation scales were set to \mtop.
A diagram removal scheme~\cite{Frixione:2008yi} was employed to handle the interference
with \ttbar production~\cite{ATL-PHYS-PUB-2016-020}.
The inclusive cross-section of $Wt$ production was corrected to the theory prediction
calculated at NLO in \alphas with NNLL soft-gluon
corrections~\cite{Kidonakis:2010ux,Kidonakis:2013zqa} yielding
\(\sigma(Wt)_{\text{NLO+NNLL}} = 71.7 \pm 3.8\,\unit{\pb}\).
The uncertainty in
the cross-section due to the PDF was calculated using the \MSTW[nnlo]
PDF set, and was added in
quadrature to the uncertainty of $\sigma(Wt)$.

The production of \ttV $(V=W, Z)$ events was modelled using the \MGNLO[2.3.3]~\cite{Alwall:2014hca} generator, which provided MEs at
NLO in \alphas
with the \NNPDF[3.0nlo] PDF set. The functional form of the renormalisation and
factorisation scales was set to
\(0.5 \times \sum_i \sqrt{m^2_i+p^2_{\text{T},i}}\),
where the sum runs over all the particles
generated from the ME calculation. Top quarks were decayed
at leading-order (LO) using \MADSPIN~\cite{Frixione:2007zp,Artoisenet:2012st}.
The events were passed through
\PYTHIA[8.210] for the PS and
hadronisation.
The cross-sections were calculated at NLO in \alphas  and NLO in $\alpha_\textrm{EW}$ accuracy using
\MGNLO as reported in Ref.~\cite{deFlorian:2016spz-update}.
In the case of $\ttbar l^+l^-$ ($l = e, \mu, \tau$), the cross-section was scaled by
an off-shell correction estimated at the one-loop level in \alphas.
The predicted values at \(\rts = \qty{13}{\TeV}\) are \(0.88^{+0.09}_{-0.11}\,\unit{\pb}\) and
\(0.60^{+0.08}_{-0.07}\,\unit{\pb}\) for $\ttbar Z$ and $\ttbar W$, respectively, where the
uncertainties were estimated from variations of \alphas and the renormalisation and
factorisation scales.

The production of $\ttbar H$ events was modelled using the
\POWHEGBOX[v2]~\cite{Frixione:2007nw,Nason:2004rx,Frixione:2007vw,Alioli:2010xd,Hartanto:2015uka}
generator.
The functional form of the renormalisation and factorisation scales was
set to \(\sqrt[3]{m_\text{T}(t)\cdot m_\text{T}(\bar{t}) \cdot m_\text{T}(H)}\),where $m_\textrm{T} = \sqrt{m^2+\pT^2}$ is the transverse mass of a generated particle, $m$ is its mass, and \pT is its transverse momentum.
The events were passed through \PYTHIA[8.230].
The cross-section was calculated at NLO in \alphas and NLO in $\alpha_\textrm{EW}$ accuracy using
\MGNLO as reported in Ref.~\cite{deFlorian:2016spz-update}.
The predicted value at \(\rts = \qty{13}{\TeV}\) is
\(507^{+35}_{-50}\,\unit{\fb}\), where the uncertainties were estimated from
variations of \alphas and the renormalisation and factorisation scales.

The background for this analysis comes from two main sources. The first is the production of $Z$ bosons decaying to pairs of leptons. The second source includes all processes in which the probe electrons do not come from the decays of real $W$ or $Z$ bosons. Such electrons are referred to as \textit{fake} electrons.

The production of most \(V+\)jets ($V = Z, W$) events
was simulated with the \SHERPA[2.2.11]~\cite{Bothmann:2019yzt}
generator. For the simulation of $Z \to \tau^+ \tau^-$ events \SHERPA[2.2.14] was used.
The simulation setup involved MEs with NLO accuracy in \alphas for up to two partons, and LO accuracy for up
to four partons calculated with the \textsc{Comix}~\cite{Gleisberg:2008fv} and
\OPENLOOPS~\cite{Buccioni:2019sur,Cascioli:2011va,Denner:2016kdg} libraries.
The default \SHERPA PS~\cite{Schumann:2007mg} based on
Catani--Seymour dipole factorisation and the cluster hadronisation model~\cite{Winter:2003tt}
were used. They employed the dedicated set of tuned parameters developed by the
\SHERPA authors and the \NNPDF[3.0nnlo]~\cite{Ball:2014uwa} PDF set.
The NLO MEs for a given jet multiplicity were matched to the PS using a colour-exact variant of the MC@NLO
algorithm~\cite{Hoeche:2011fd}. Different jet multiplicities were then merged
into an inclusive sample using an improved CKKW matching
procedure~\cite{Catani:2001cc,Hoeche:2009rj} which was extended to NLO
accuracy using the \MEPSatNLO prescription~\cite{Hoeche:2012yf}.

The production of diboson final states (\(VV\)) was simulated with the
\SHERPA[2.2.1] or 2.2.2 generator depending on the process,
including off-shell effects and Higgs boson contributions, where appropriate.
Fully leptonic final states and semileptonic final states, where one boson
decays leptonically and the other hadronically, were generated using
MEs at NLO in \alphas for up to one additional parton
and at LO in \alphas for up to three additional parton
emissions. Samples for the loop-induced processes \(gg \to VV\) were
generated using LO-accurate MEs for up to one
additional parton emission for both the cases of fully leptonic and
semileptonic final states. The ME calculations were matched
and merged with the \SHERPA PS based on Catani--Seymour
dipole factorisation~\cite{Gleisberg:2008fv,Schumann:2007mg} using the MEPS@NLO
prescription~\cite{Hoeche:2011fd,Hoeche:2012yf,Catani:2001cc,Hoeche:2009rj}.
The virtual QCD corrections were provided by the
\OPENLOOPS library~\cite{Buccioni:2019sur,Cascioli:2011va,Denner:2016kdg}. The
\NNPDF[3.0nnlo] set of PDFs was used, along with the
dedicated set of tuned PS parameters developed by the
\SHERPA authors.

Events featuring at least one fake lepton
originate from $\ttbar$, $\ttbar+X$ ($X = W, Z, H$), or $W t$ production processes.
The simulation of such events
was described above. Single-top-quark $s$- and $t$-channel production can also produce dilepton
events with at least one fake lepton.
These processes were modelled using the
\POWHEGBOX[v2]~\cite{Frederix:2012dh,Nason:2004rx,Frixione:2007vw,Alioli:2010xd}
generator.
The functional form of the
renormalisation and factorisation scales was set to
\(\sqrt{m_b^2 + \pTX[2][b]}\) for $t$-channel following the
recommendation of Ref.~\cite{Frederix:2012dh}, and to \mtop for $s$-channel.
The events were processed through
\PYTHIA[8.230].
Any possible contribution of other sources of fake leptons was taken into account in the correction factors measured using data as explained in Section~\ref{sec:background}.

The uncertainties related to the modelling of the signal and background production were estimated by comparing the baseline MC samples described above with different MC generators or a modified configuration of the baseline MC generators. More details of these studies are given in Section~\ref{sec:fit}. Here, the description of the alternative MC samples %
is given.

The uncertainty due to the choice of the PS and hadronisation model used in \ttbar and $Wt$ production was evaluated with a sample generated with \POWHEGBOX[v2] for the hard-scatter simulation but passed through \HERWIG[7.1.3] for the PS and hadronisation. The samples employ the \HERWIG[7.1] default set of tuned parameters and the \MMHT[lo]~\cite{Harland-Lang:2014zoa} PDF set.

The baseline \ttbar and $Wt$ samples were generated with \hdamp = 1.5~\mtop.
A variation of the \hdamp value to 3.0~\mtop was used to evaluate the systematic uncertainty related to this parameter.

The NLO matching uncertainty
was evaluated by comparing the baseline \ttbar and $Wt$ samples with alternative samples obtained by setting the \texttt{pT$_{\textrm{hard}}$} parameter of \PYTHIA[8] to one (the default value is zero). This parameter regulates the definition of the vetoed region of the showering, important to avoid holes or overlap in the phase space filled by \POWHEG and \PYTHIA. This follows the recommendation of Refs. \cite{Hoche:2021mkv,ATL-PHYS-PUB-2023-029}.

An additional source of uncertainty in the PS is the modelling of the recoil for gluons emitted from $b$-quarks in the $t \to W b$ process.  The baseline \ttbar sample employs a scheme where partons recoil against $b$-quarks. An alternative sample was generated with identical settings to the baseline ones, except for a modified recoil scheme. Here, the gluon emissions from the PS of the $b$-quark were made to recoil against the top quark itself. This scheme changes the modelling of second and subsequent gluon emissions.

The baseline \POWPY[8] sample of $Wt$ production was generated using the diagram removal scheme to model the interference between the \ttbar and $Wt$ production. To evaluate the corresponding uncertainty, the baseline simulation was compared with an alternative sample generated using the diagram subtraction scheme~\cite{Frixione:2008yi,ATL-PHYS-PUB-2016-020}.

To estimate the systematic uncertainty related to the electron reconstruction efficiency, an alternative sample of $Z( \to e^+ e^-) +$ jets events was used,
which was simulated with \MGNLO[2.2.2]~\cite{Alwall:2014hca}, using MEs at NLO in \alphas with up to four final-state partons. The ME calculation employed the \NNPDF[2.3lo] set of PDFs. Events were passed through \PYTHIA[8.186] for the modelling of the PS and hadronisation. The overlap between ME and PS emissions was removed using the CKKW-L merging procedure~\cite{Lonnblad:2001iq,Lonnblad:2011xx}.
The \(Z\)+jets samples were normalised to a NNLO prediction~\cite{Anastasiou:2003ds}.

%
%
%
%
%

%
%

%
%
%
%
%
%
%
%
%
%
%
%
%
%

%

%
%
%
%
%
%
%

%

% End of text imported from the .//sections/04_Data_and_Simulation.tex input file

% The next lines are included from the .//sections/05_Event_Selection.tex input file
%
%
%
%
%
%
%
%
%
%
%
%

\section{Event reconstruction and selection}
\label{sec:events}

A sample of \ttbar events, in which the $W$ bosons from both $t$ and $\bar t$ quarks decay to leptons, is selected for the measurement. With this selection, the tag-and-probe approach is used. One charged lepton (muon or electron) from the decay of the $t\bar{t}$ pair is used as the \textit{tag}. The other lepton (electron), which is called the \textit{probe}, is used in an unbiased way to determine whether the electron is directly produced in the decay $W\rightarrow e\nu_e$ or via an intermediate decay $W\rightarrow\tau \nu_{\tau} \to e \nu_e \nu_{\tau} \nu_{\tau} $. In the first case, the probe electrons are called \textit{prompt} and denoted by \Wtoe, while in the second case they are denoted by \Wtotaue.

To distinguish between electrons from \Wtoe and \Wtotaue decays, the analysis exploits
differences in the \pT spectra of the probe electrons together with the displacement of the intermediate $\tau$-lepton decays from the $pp$ collision point (primary vertex) owing to the $\tau$-lepton lifetime. The quantity reflecting the displacement of the vertex of the $\tau$-lepton decay from the primary vertex is the transverse impact parameter (\dzero) of the electron track with respect to the beamline. This quantity is defined as the distance of closest approach of the extrapolated track to the beamline in the transverse plane. Electrons originating from the $W\rightarrow\tau \nu_{\tau} \to e \nu_e \nu_{\tau} \nu_{\tau}$ decays generally have lower \pT due to energy lost to undetected neutrinos. These electrons also have on average a larger $|\dzero|$ because of the non-zero lifetime of the $\tau$-lepton. In contrast, the electrons from prompt decays are produced directly at the primary vertex, and therefore their measured value of \dzero depends mainly on the detector resolution.

Electrons are reconstructed using the clusters of energy deposits in the electromagnetic calorimeter matched to tracks in the ID. They must pass the \textit{tight} (for both isolation and
quality) requirements of
Refs.~\cite{ATLAS:2019qmc, ATLAS:2023dxj}.
Muons are reconstructed by combining tracks from the ID with matching tracks reconstructed in the MS. They must pass the \textit{medium} quality and \textit{tight} isolation criteria defined in Ref.~\cite{ATLAS:2020auj}. The charged tracks of the selected electrons and muons must be consistent with the primary vertex and have $|\dzero| < 0.5$~mm. These requirements reduce the contribution of fake leptons, which becomes large compared to the signal for $|\dzero| > 0.5$~mm. The pseudorapidity of the electromagnetic cluster associated with the electron ($\eta_{\textrm{cl}}$) must be $|\eta_{\textrm{cl}}| < 2.47$. The electrons with $ 1.37 < |\eta_{\textrm{cl}}| < 1.52$ are not used in the analysis to exclude the transition region between the barrel and endcap electromagnetic calorimeters.
The pseudorapidity of the selected muons must be $|\eta| < 2.5$.
All these requirements are applied to both tag and probe leptons. The tag lepton must pass a suite of single-muon or single-electron triggers \cite{TRIG-2019-03,ATLAS:2019dpa,ATLAS:2020gty}. No trigger requirement for the probe electron is applied. The lowest electron \pT trigger threshold is 24 GeV in 2015 and 26 GeV in 2016--2018. The lowest muon \pT trigger threshold is 20 GeV in 2015 and 26 GeV in 2016--2018. The efficiency of the triggers reached the plateau for electrons with $\pT > 27$ GeV and muons with $\pT > 27.3$ GeV. Consequently, the tag electrons must have $\pT > 27$ GeV and tag muons must have $\pT > 27.3$ GeV.
The probe electrons must have $7 < \pT < 250$~GeV. The fraction of electrons with $\pT > 250$~GeV from $W \to \tau \nu_\tau$ is much smaller than from other sources, and the upper cut on \pT of probe electrons is applied to simplify the measurement without loss in accuracy.

Jets are reconstructed using the anti-$k_t$ algorithm \cite{Cacciari:2008gp,Cacciari:2011ma} with a radius parameter R = 0.4 applied to particle flow objects based on tracks and topological clusters calibrated %
as specified in Ref. \cite{PERF-2015-09}. The jet energy scale and resolution are calibrated using simulations with in situ corrections obtained from data \cite{ATLAS:2020cli}.
Jets are required to have $\pT > 25$~GeV and $|\eta| < 2.5$. To reduce the \pileup contribution, an additional requirement for jets with $\pT < 60$ GeV and $|\eta| < 2.4$ is applied based on a dedicated jet vertex tagger (JVT) algorithm which is designed to discriminate between jets produced in the hard-scatter process and those from \pileup~\cite{ATLAS:2015ull}.

To avoid double counting electrons and muons within jets, an overlap removal procedure is implemented.
Specifically, all electrons that share a track with another electron are removed,
as are
electrons sharing a track with a muon. Additionally,
the closest jet found within a $\Delta R=0.2$ of a reconstructed electron is removed.
Any electron subsequently found within $\Delta R=0.4$ of a jet is also removed.
Any jet with fewer than three tracks associated to it and found within $\Delta R=0.2$ of a muon is removed.
Any muon subsequently found within $\Delta R = 0.4$ of a jet is removed.

Jets likely to contain $b$ hadrons are identified as $b$-jets using the \texttt{DL1r} algorithm \cite{ATLAS:2022qxm}, which employs a deep neural network.
Such jets are referred to as \textit{$b$-tagged} in the following. The network utilises distinctive features of $b$-hadrons, such as the impact parameters of tracks and the displaced vertices reconstructed in the ID. The $b$-tagging efficiency in simulated \ttbar events is approximately 70\%.

Events containing exactly one tag lepton (muon or electron), one probe electron, and at least two $b$-tagged jets are selected for this measurement.
The tag and probe leptons must have opposite electric charge. The events with $\mu_\textrm{tag}$ and $e_\textrm{probe}$, and with $e_\textrm{tag}$  and $e_\textrm{probe}$ are analysed separately. These channels are referred to as $\mu e$ and $e e$, respectively.
To further refine the selection, several criteria based on the invariant mass  of the tag and probe leptons ($m_{\ell \ell}$) are applied.
Only events with  $m_{\ell \ell} > 15$ GeV are selected to suppress the contribution from fake probe electrons.
Additionally, the $ee$ channel events with $|m_{e^+e^-} - m_{Z}| < 5$ GeV are excluded, to remove events that could come from $Z$-boson production.\footnote{The mass of $Z$ boson is taken to be $m_{Z} = 91.188$ GeV \cite{Workman:2022ynf}.}
Finally, the $ee$ events where both electrons satisfied the tag criteria are analysed twice, with each electron considered in turn the tag or probe.

The sample of events passing this selection is referred to as \textit{nominal} in the following.

% End of text imported from the .//sections/05_Event_Selection.tex input file

% The next lines are included from the .//sections/06_Calibration.tex input file

\section{Electron impact parameter calibration}
\label{sec:d0_cal}

The impact parameter \dzero is one of the two quantities used to separate the electrons produced in $\tau$-lepton decays from other sources. An accurate calibration of the \dzero distribution in simulation is essential for the measurement of \Rtaue. The impact parameter \dzero is measured relative to the %
beamline.
Therefore, the shape of the \dzero distribution of prompt electrons is mainly determined by the detector resolution and the precision of the beamspot measurement,
and can be obtained in data using the $Z \to e^+ e^-$ decay. A clean sample of such events is obtained by requiring that two electrons
satisfy the same criteria as in the nominal event selection, but
with the invariant mass of the $e^+e^-$ pair satisfying $|m_{e^+e^-} - m_{Z}| < 10$ GeV. At least two jets are required in the event to reproduce the track activity around the electron as in the nominal sample. However, no $b$-tagging condition must be fulfilled. This selection results in a sample of 4.3 million electrons, which is referred to in the following as the $Z$ data sample. The contribution of non-prompt electrons is estimated by using the MC simulation to be about 0.4\%. By using the $Z$ data sample to calibrate the \dzero distribution of prompt electrons, the uncertainties due to the shape of the \dzero distribution are considerably reduced.

\begin{figure}[tb]
\centering
\begin{center}
\subfloat[]{\includegraphics[width=0.49\textwidth]{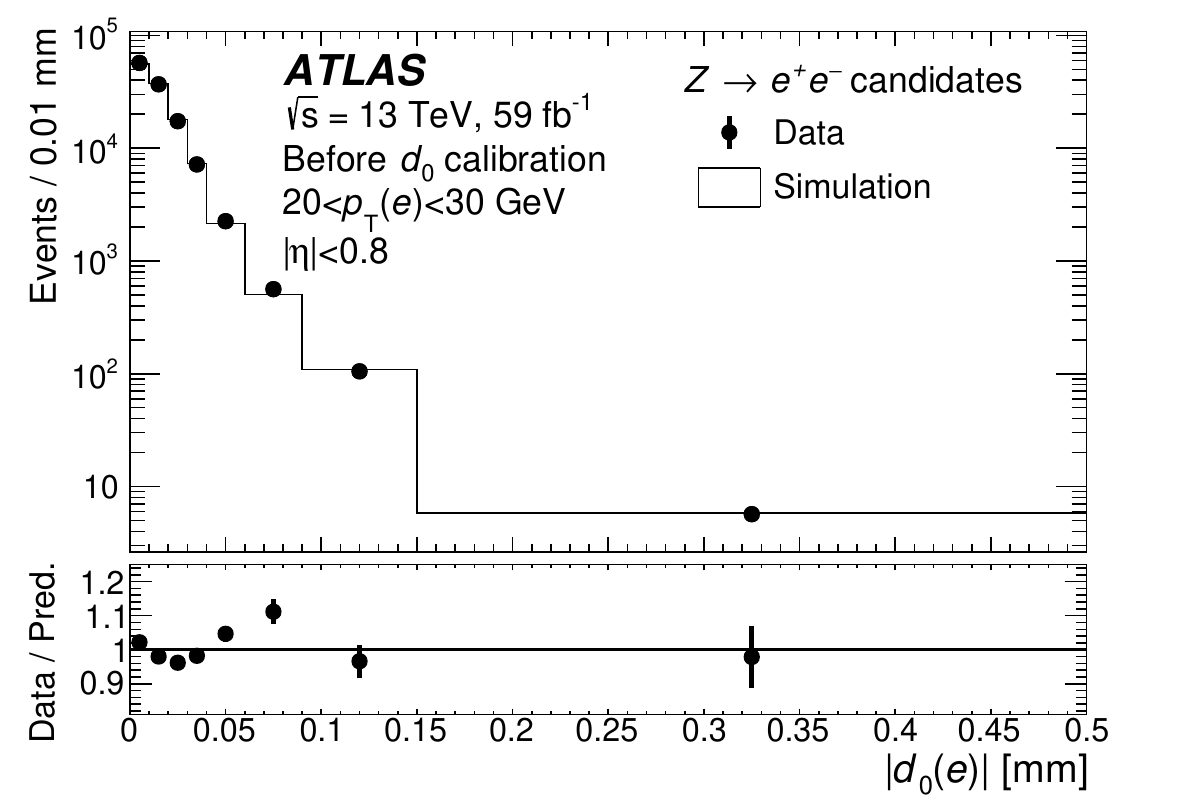}}
\subfloat[]{\includegraphics[width=0.49\textwidth]{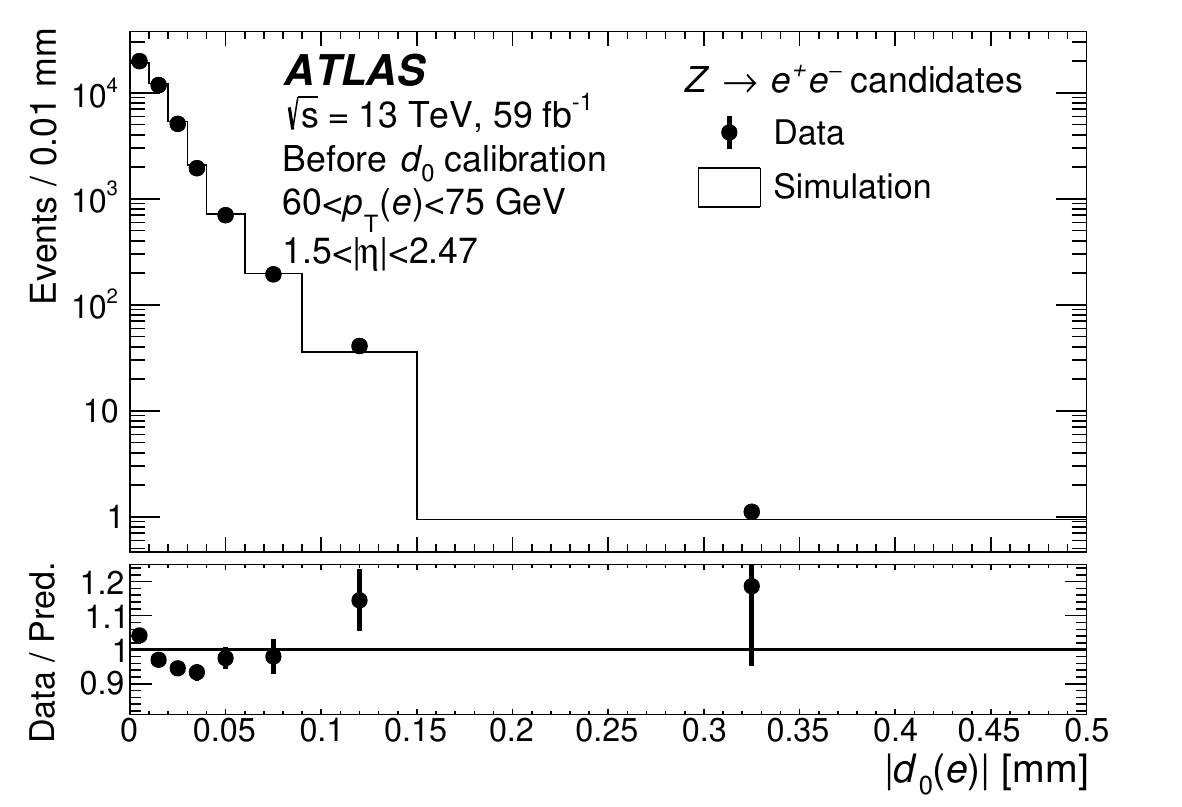}}
\end{center}
\caption{
Example distributions of $|\dzero|$ of the probe electrons for two kinematic bins in the $Z$ data and simulation samples (before $|\dzero|$ calibration) collected in the 2018 data-taking period. The data are represented by black markers and the simulation prediction as the histogram. Only the statistical uncertainties in the data are shown. The distribution of the simulated events in each kinematic bin is normalised to the number of data events in that bin. The bottom panels show the ratio of the data to the simulation prediction.
}
\label{fig:d0bins}
\end{figure}

The \dzero distribution depends on the electron \pT and $\eta$. Consequently, the shape of the \dzero distribution of prompt electrons is determined separately in 39 kinematic bins of $\pT$ and $|\eta|$. These bins are obtained by dividing the $\pT$ range of 7--250 GeV into 13 bins and the $|\eta|$ range of 0--2.47 into three bins.
The boundaries are optimised to have reasonable statistics in each bin.
Figure~\ref{fig:d0bins} shows the comparison of $|\dzero|$ distributions in data and MC simulation in two kinematic bins before the $|\dzero|$ calibration is applied.

For the measurement of \Rtaue, the \dzero distribution in each kinematic bin is taken from the $Z$ data sample after subtracting the contribution of non-prompt electrons estimated from MC simulation. The resulting distributions $T_{i j}^{\textrm{pr}}(\dzero)$, which are referred to as \textit{\dzero templates},
are normalised to unity. Here the indices $i$ and $j$ refer to the \pT and $|\eta|$ bins, respectively.
Separate templates are used for data collected in different years (2015--2016, 2017, and 2018) because of the variation of the beam conditions and the alignment of the ID from year to year.

\begin{figure}[tb]
\centering
\subfloat[]{\includegraphics[width=0.42\textwidth]{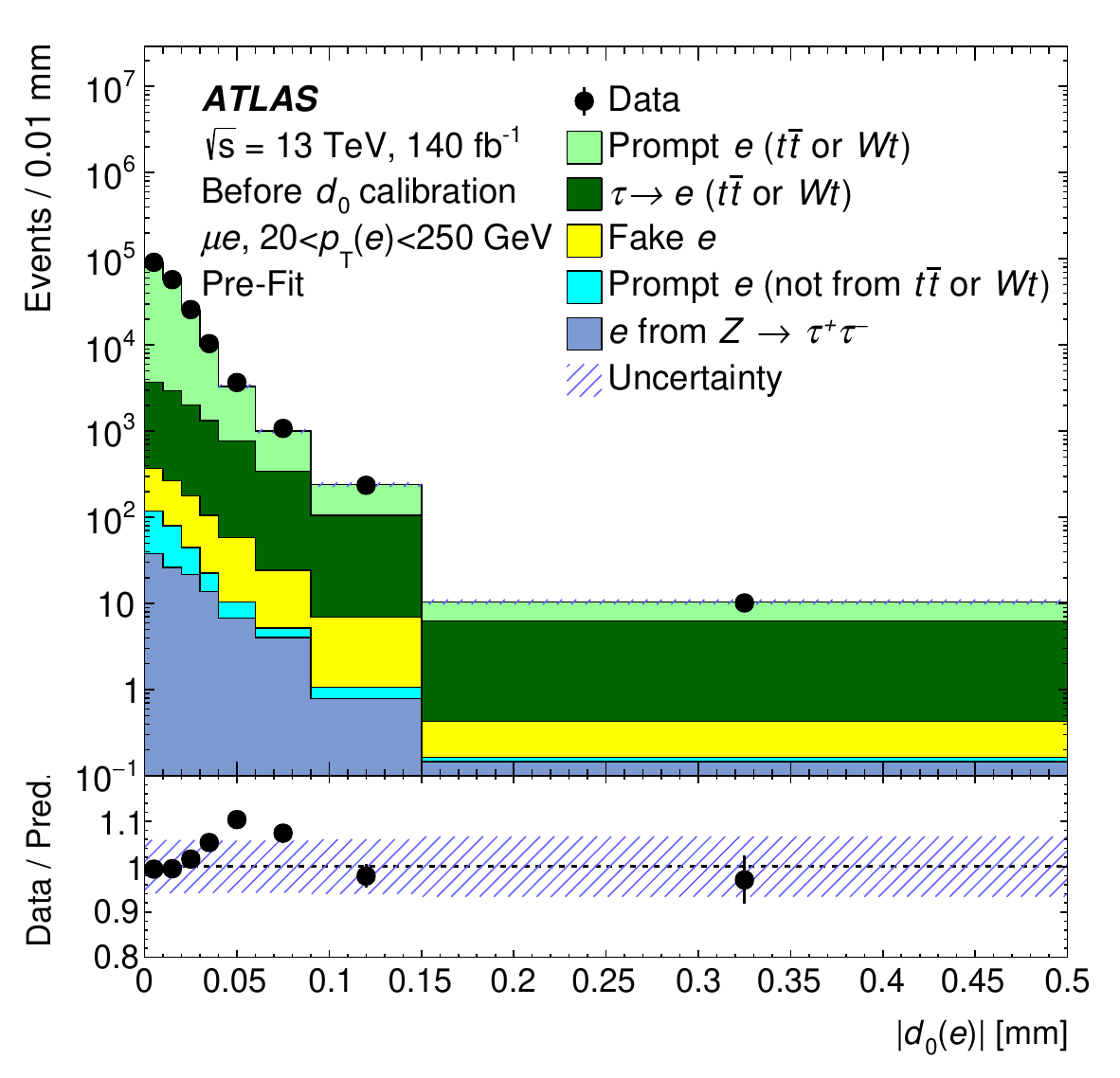}}
\subfloat[]{\includegraphics[width=0.42\textwidth]{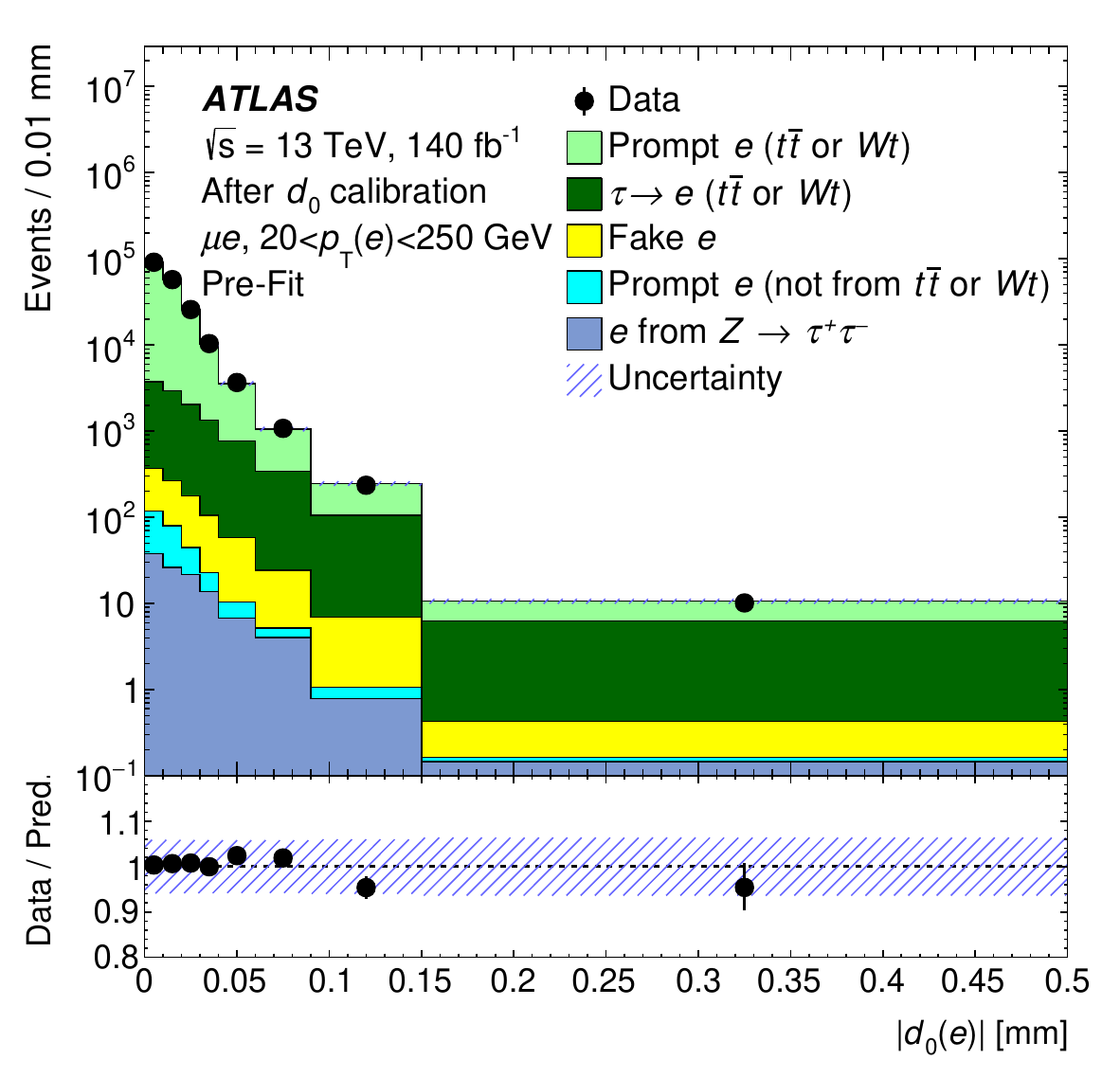}}
\caption{Distributions of $|\dzero|$ for the $\mu e$ channel of the nominal sample for $20<\pT(e)<250$ GeV (a) before and (b) after the \dzero calibration is applied. The data are represented by black markers. The different contributions to the nominal sample,
are shown by stacked histograms. The bottom panels show the ratio of the data to the simulation prediction. The systematic uncertainties in the prediction are represented by the blue hatched bands.
}
\label{fig:prefitd0_me}
\end{figure}

The \dzero distribution of prompt electrons with \pT in a range $\pT^{\textrm{min}} \le \pT < \pT^{\textrm{max}}$
is computed as
\begin{equation}
D^{\textrm{pr}}(\dzero,\pT^{\textrm{min}}, \pT^{\textrm{max}}) = \sum_{i=i_{\textrm{min}}}^{i_{\textrm{max}}} \sum_{j=1}^{3} f_{ij}^{\textrm{pr}}\, T_{ij}^{\textrm{pr}}(\dzero).
\end{equation}
Here, $f_{ij}^{\textrm{pr}}$ is the fraction of prompt probe electrons produced in the \Wtoe decay and contained in the kinematic bin $ij$ relative to the total number of prompt probe electrons from the \Wtoe decay in all bins.
The outer sum ranges from $i_{\textrm{min}}$ to $i_{\textrm{max}}$, the indices of the bins containing $\pT^{\textrm{min}}$ and $\pT^{\textrm{max}}$. The values of $\pT^{\textrm{min}}$ and $\pT^{\textrm{max}}$ correspond to the definition of \pT bins used for the measurement of \Rtaue ($7 < \pT < 10$ GeV, $10 < \pT < 20$ GeV, and $20 < \pT < 250$ GeV) as defined in Section~\ref{sec:fit}.
The same method and \dzero templates are used to obtain the \dzero distribution of prompt electrons from other sources (i.e., non-resonant Drell-Yan production of $e^+e^-$ and $VV$ production), with the fractions $f_{ij}$ being computed from MC simulation for each source separately. Figure \ref{fig:prefitd0_me} shows the comparison of $|\dzero|$ distributions of the probe electrons with $20 < \pT < 250$ GeV in the nominal sample before and after the \dzero calibration is applied. A significant improvement in the agreement between data and MC simulation is observed.

The \dzero templates of electrons from \Wtotaue decays are determined using simulated events and corrected for the difference in resolution of \dzero in data and simulation following the method employed in Ref.~\cite{ATLAS:2020xea}.
For these electrons,
the \dzero distribution is the convolution of the \dzero resolution and the \dzero distribution due to the lifetime of the $\tau$-lepton. This  lifetime is known with good precision \cite{Belle:2013teo}.
Therefore,
only the correction of \dzero resolution is necessary to calibrate the \dzero templates
of electrons from \Wtotaue decays in simulation.
To quantify the difference in resolution of \dzero in data and simulation, the $|\dzero|$ distribution of the probe electrons in the $Z$ data sample is fitted using a Gaussian function with the mean fixed at zero. The fit is performed in each kinematic bin in the range of $|\dzero| < 0.02$~mm. For $\pT = 20$ GeV, the \dzero resolution is about 15 $\mu$m.
The \dzero resolution is determined separately for 2015-2016, 2017, and 2018 data-taking periods.
The \dzero templates of the electrons from $\tau$-lepton decays in the MC simulation are corrected using the measured $|\dzero|$ resolution.
The corrected templates are then used to measure \Rtaue.
The resulting correction due to the  \dzero resolution is small and varies within 1\% for different values of \dzero.
The \dzero templates of fake electrons are similarly taken from simulated events and corrected using the \dzero resolution from data events
with corrections also at the 1\% level.

\FloatBarrier

% End of text imported from the .//sections/06_Calibration.tex input file

% The next lines are included from the .//sections/07_Background.tex input file
%
%
%
%
%
%
%
%
%
%
%
%
%
%
%
%
%

\section{Background estimation}
\label{sec:background}

\begin{figure}[tb]
\centering
\includegraphics[width =0.6\textwidth]{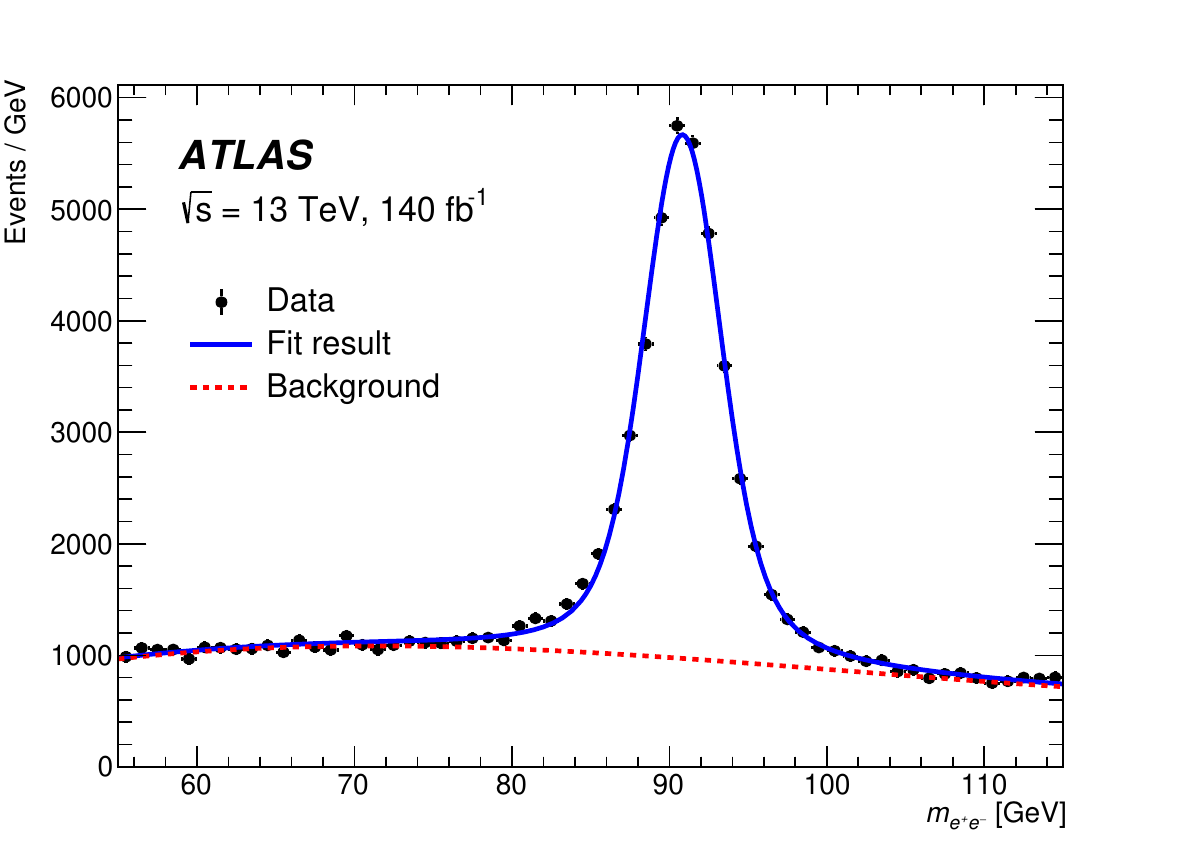}

\caption{Distribution of the invariant mass of the electron-positron system, $m_{e^+e^-}$, in the control sample. The data are represented by black markers. The result of the fit by the sum of the Voigt profile for the $Z$-boson peak and the second-degree Chebyshev polynomial for the background is shown by the solid line. The dashed line corresponds to the parametrisation of the background contribution.}
\label{fig:z0eefit}

\end{figure}

Drell-Yan production of an $e^+e^-$ pair in association with jets is a large background at small values of $|\dzero|$. The resonant $Z \rightarrow e^+e^-$ decay is suppressed by applying a selection on the $e^+e^-$ invariant mass $|m_{e^+e^-} - m_{Z}| > 5$ GeV. However, non-resonant  $e^+e^-$ production
is an irreducible background. Its contribution
is determined using a control sample of $e^+e^-$ events containing the resonant $Z \rightarrow e^+e^-$ contribution. This sample is obtained by removing the $Z$-mass window veto ($|m_{e^+e^-} - m_{Z}| > 5$ GeV), whilst keeping all other requirements as specified in Section~\ref{sec:events}. Figure \ref{fig:z0eefit} shows the $m_{e^+e^-}$ distribution for the data control sample in the range of $55 < m_{e^+e^-} < 115$~GeV which is used to determine the  $Z \rightarrow e^+e^-$ contribution.

The control sample in data and MC simulation is partitioned into three subsamples, depending on \pT of the probe electron ($7 < \pT < 10$ GeV, $10 < \pT < 20$ GeV, and $20 < \pT < 250$ GeV). This binning in \pT matches the one used in the measurement of \Rtaue, as detailed in Section~\ref{sec:fit}.
A binned extended maximum-likelihood fit \cite{Barlow:1990vc} of the $m_{e^+e^-}$ distribution in the range of $55 < m_{e^+e^-} < 115$~GeV is used to estimate the number of $Z$+jets events in each \pT bin $i$, $N_{Z,i}$, of the control sample. The distribution is modelled by the function
\begin{equation}
F(m_{e^+e^-} |\,m_Z,\Gamma_Z,s,N_{Z,i},N_{\textrm{bkg},i}) = N_{Z,i} \, V(m_{e^+e^-} | \,m_Z,\Gamma_Z,s) + N_{\textrm{bkg},i} \, P(m_{e^+e^-}).
\label{eq:Z0}
\end{equation}
Here, $V(m_{e^+e^-} | \,m_Z,\Gamma_Z,s)$ is the Voigt profile \cite{Voigt1912} describing the $Z \rightarrow e^+e^-$ signal and $P(m_{e^+e^-})$ is the second-degree Chebyshev polynomial corresponding to the background under the $Z$-boson peak. The parameters $N_{Z,i}$ and $N_{\textrm{bkg},i}$ are the numbers of $Z$ and background events, respectively. The parameters $m_Z$ and $\Gamma_Z$ are the mass and width of the Z boson, respectively, and $s$ is the standard deviation of the mass resolution. The $Z$ boson's width is fixed to the PDG value $\Gamma_Z = 2.4955$ GeV \cite{Workman:2022ynf}, while the other parameters in Eq.~(\ref{eq:Z0}) are varied freely in the fit.

The numbers of $Z \rightarrow e^+e^-$ events in data and MC, $N_{\textrm{Z},i}^{\textrm{data}}$ and $N_{\textrm{Z},i}^{\textrm{MC}}$, are obtained from the fit to the $m_{e^+e^-}$ distribution in data and MC simulation, respectively.
The simulated non-resonant Drell-Yan contribution in the nominal sample is then scaled in each \pT bin $i$ by the factor $C^Z_i \equiv N_{\textrm{Z},i}^{\textrm{data}}/N_{\textrm{Z,i}}^{\textrm{MC}}$. The scaling factors $C^Z_i$ are also applied to the $Z(\to \tau^+ \tau^-)+$jets sample, which contributes to both $\mu e$ and $e e$ channels.

The values of $C^Z_i$ in all \pT groups are found to be statistically consistent across all \pT bins. The average value of $C^Z_i$ is $1.117 \pm 0.013$, where the uncertainty reflects the limited size of the data and simulation samples.

The solid blue line in Figure \ref{fig:z0eefit} shows the result of the fit to the full data control sample using the model function specified in Eq. (\ref{eq:Z0}). The parametrisation of the background, which comes mainly from \ttbar and $Wt$ production, is shown as the dashed line.

Fake electrons are another important source of background. They mainly
arise from two processes: the decay of $b$- or $c$-hadrons, and photon conversions.
The number of fake electrons in the nominal sample in simulation is corrected by a factor derived from the comparison of data and simulation. A control sample is defined by applying the nominal selections, but requiring two leptons with the same electric charge. The fake electrons contribute to both the nominal (i.e., opposite-sign) and control (i.e, same-sign) samples, while the contribution of non-fake leptons to the same-sign sample is reduced.
For electrons with $\pT < 10$ GeV the fraction of fake electrons in the same-sign sample is about 97\%, and for electrons with $20 < \pT < 250$ GeV it is about 50\%.
The extrapolation of the correction factor from the same-sign to opposite-sign sample is evaluated from simulated events.

In addition to the fake electrons from hadron decays and photon conversions, prompt electrons from $W$- and $Z$-boson decays also contribute to the same-sign sample, especially at high $\pT$ and small values of $|\dzero|$. The main processes yielding prompt same-sign electrons are $\ttbar+V$ and $VV$ production, where the leptons from the $V$ boson and one of the top-quark decays can have the same electric charge.
There is also a 3\% contribution of events with a wrong measurement of the charge of the electron due to bremsstrahlung.
The correction factors for the number of fake electrons from hadron decays ($C^\mathrm{HAD}$), from photon conversions ($C^\mathrm{PH}$), and for the number of prompt electrons ($C^\mathrm{PR}$), are derived from the comparison of the number of same-sign events as a function of probe lepton \pT in data and simulation. In addition to the events with probe electrons, the same-sign events with probe muons are also used. The probe muons are selected using the same requirements as the tag muons but changing the \pT condition to $5 < \pT(\mu) < 250$ GeV. No trigger requirements are applied to the probe muons.  The addition of the events with a probe muon increases the statistical precision of the correction factors.
The same $C^\mathrm{PR}$ value is applied to the events with prompt probe muons and electrons, and the same $C^\mathrm{HAD}$ value is applied to the events with fake muons and electrons from hadron decays. It was verified that the separate treatment of electrons and muons from these sources had no statistically significant impact on the result.

\begin{figure}[tb]
\centering
\subfloat[]{\includegraphics[width=0.49\textwidth]{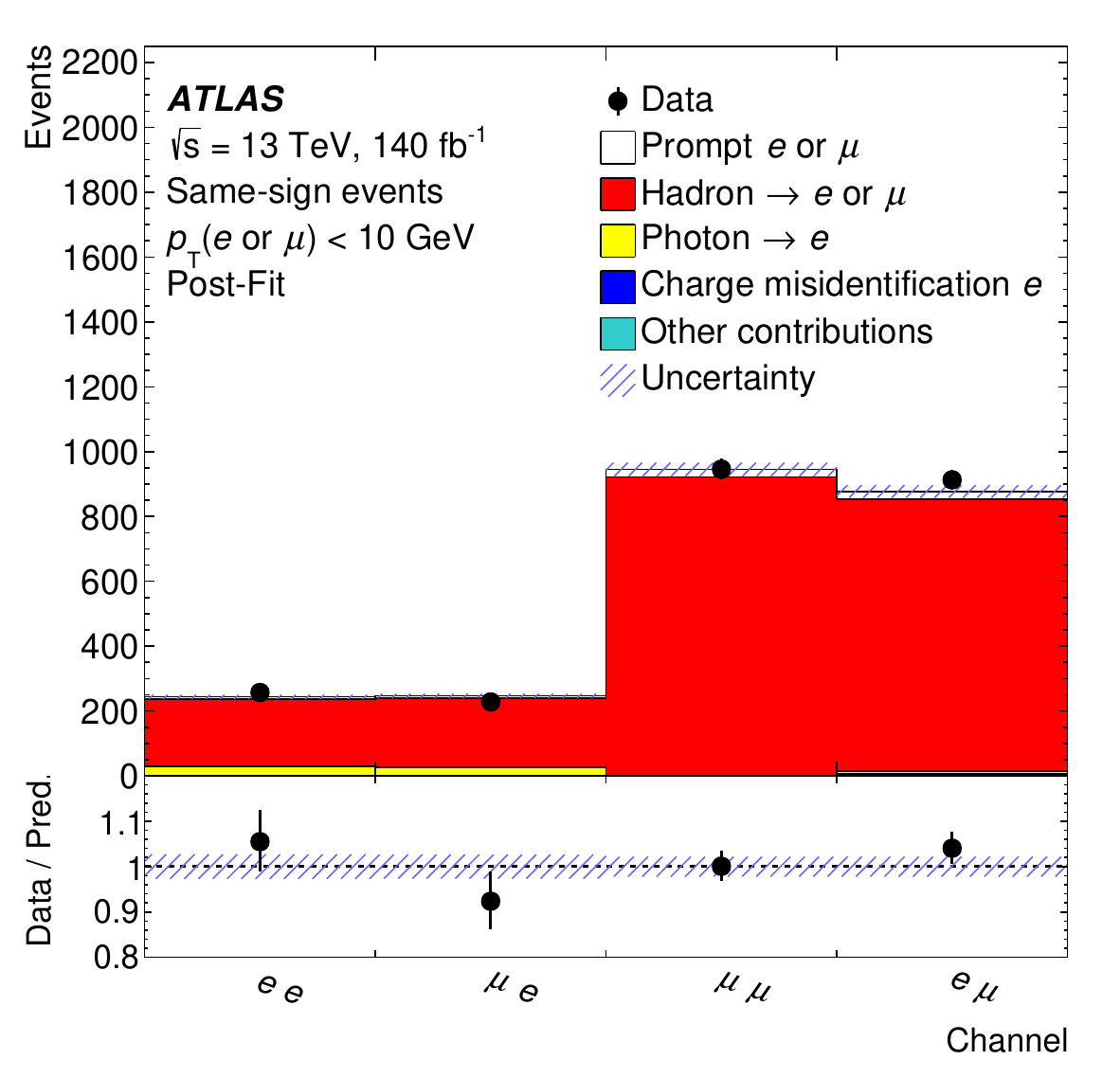}}
\subfloat[]{\includegraphics[width=0.49\textwidth]{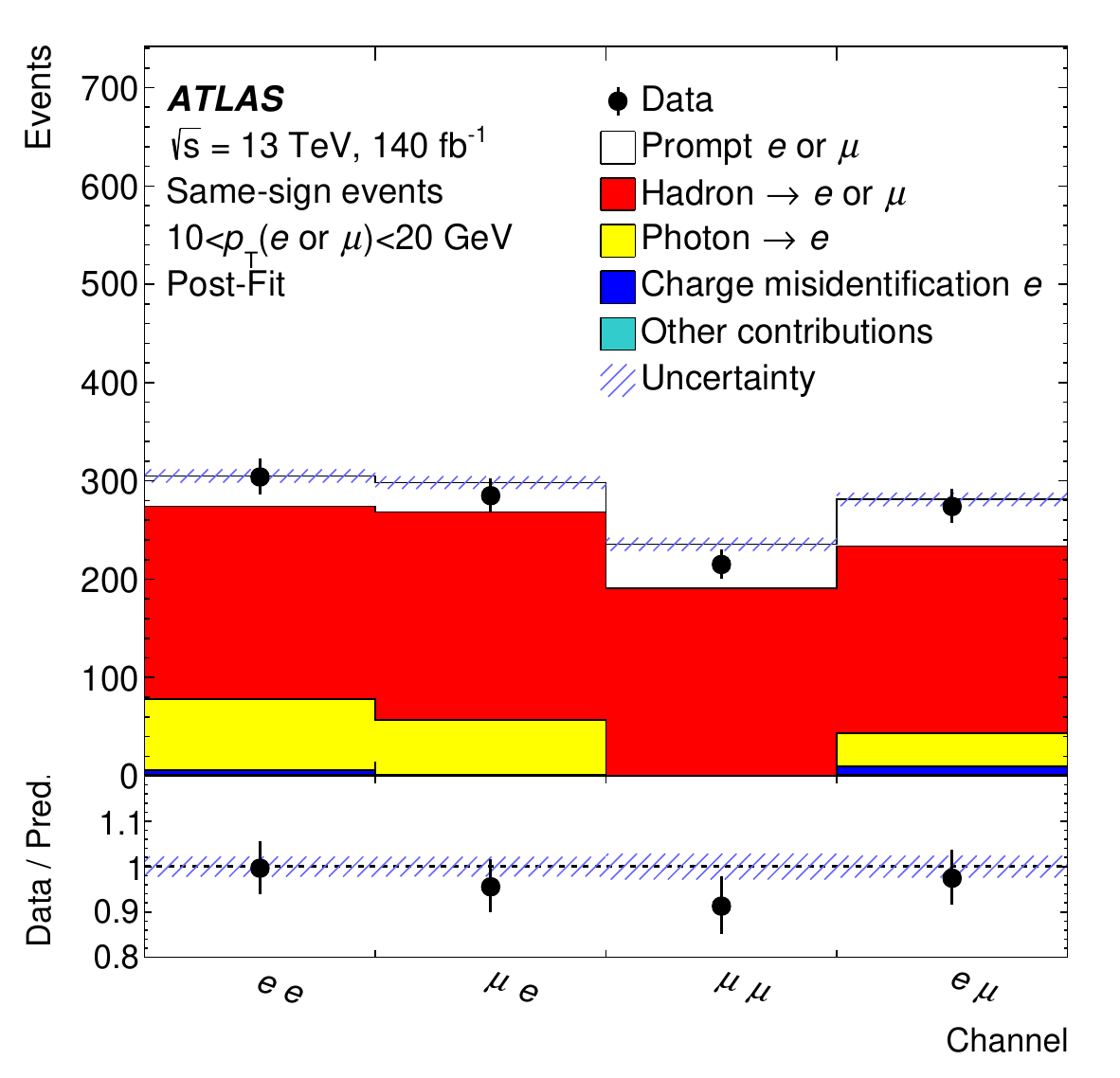}} \\
\subfloat[]{\includegraphics[width=0.49\textwidth]{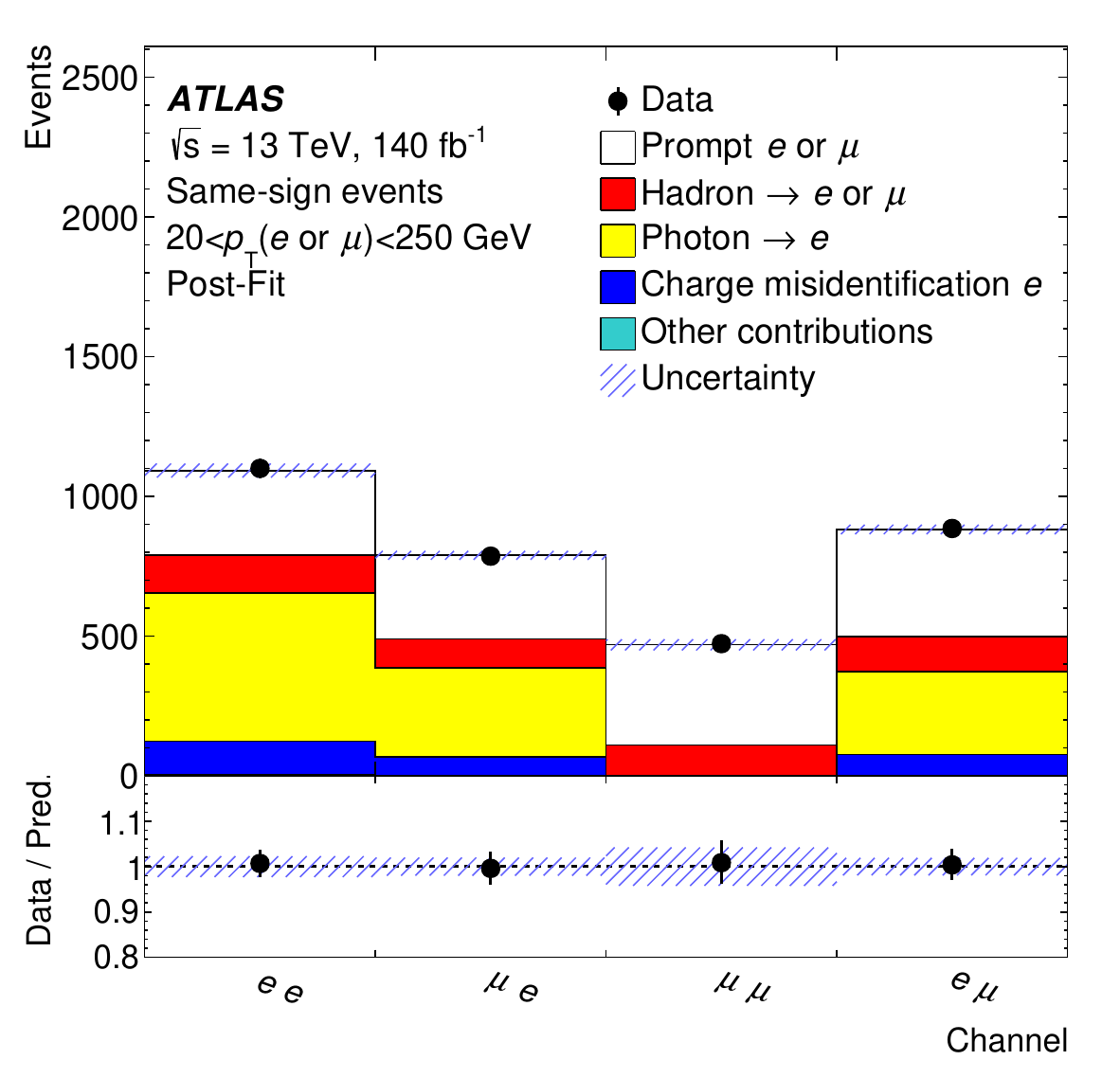}}
\caption{Number of same-sign events in the $ee$, $\mu e$, $e \mu$, and $\mu \mu$ channels where the first lepton is the tag lepton and the second is the probe lepton. The sub-figures correspond to (a) $\pT< 10$ GeV, (b) $10 < \pT < 20$ GeV, and (c) $20 < \pT < 250$ GeV of the probe lepton (electron or muon). The data are represented by black markers and the different components contributing to the sample, taken from simulation after the fit (`Post-Fit'), are shown by stacked histograms. The component `Other contributions' contains events with both fake leptons or with charge misidentification of both leptons. Distributions are shown after the fit to the same-sign data has been performed. The systematic uncertainties related to the limited size of the simulation samples are included in the fit, and represented by the blue hatched bands. The bottom panels show  the ratio of the data to the simulation prediction.
}
\label{fig:fakesFit}
\end{figure}

The same-sign sample is divided into 12 groups depending on the channel ($ee$, $\mu e$, $e \mu$, and $\mu \mu$) and \pT of the probe lepton ($\pT < 10$ GeV, $10 < \pT < 20$ GeV, and $20 < \pT < 250$ GeV), and a binned maximum-likelihood fit is performed to these groups. The free parameters in the fit are $C^\mathrm{HAD}$, $C^\mathrm{PH}$, and $C^\mathrm{PR}$. The statistical and systematic uncertainties related to the limited size of the simulation samples are taken into account. Other systematic uncertainties are discussed in Section~\ref{sec:fit}.
Figure \ref{fig:fakesFit} shows the results of the fit.
A good agreement between data and simulation is obtained for all 12 groups of same-sign events with the $p$-value of the global goodness of the fit \cite{cousins}
being equal to 0.7.
The obtained values of $C^\mathrm{HAD}$, $C^\mathrm{PH}$, and $C^\mathrm{PR}$ are $1.41 \pm 0.03$, $0.84 \pm 0.05$, and $1.45 \pm 0.08$, respectively.

\begin{figure}[tb]
\begin{center}
\subfloat[]{\includegraphics[width=0.49\textwidth]{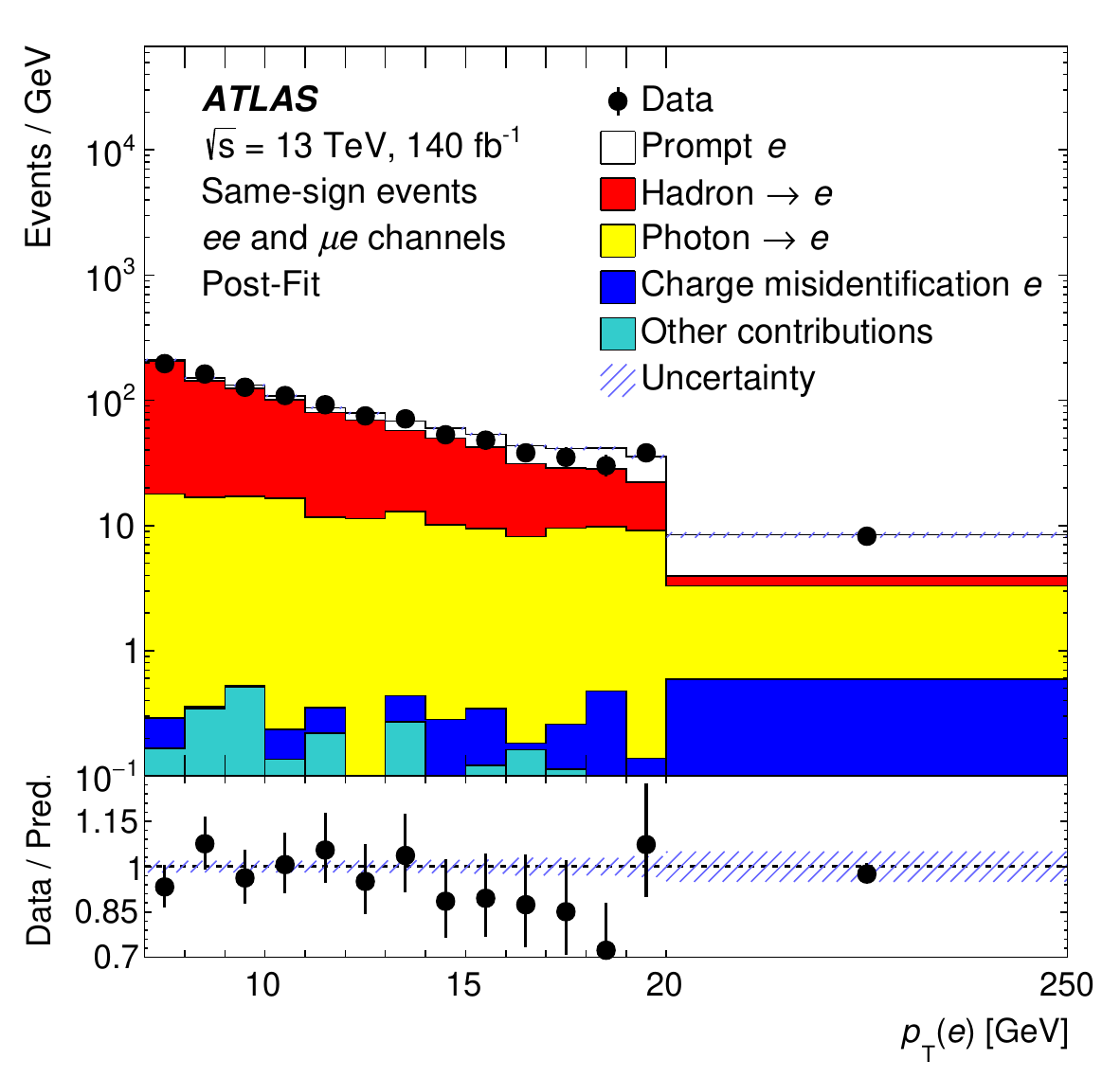}}
\subfloat[]{\includegraphics[width=0.49\textwidth]{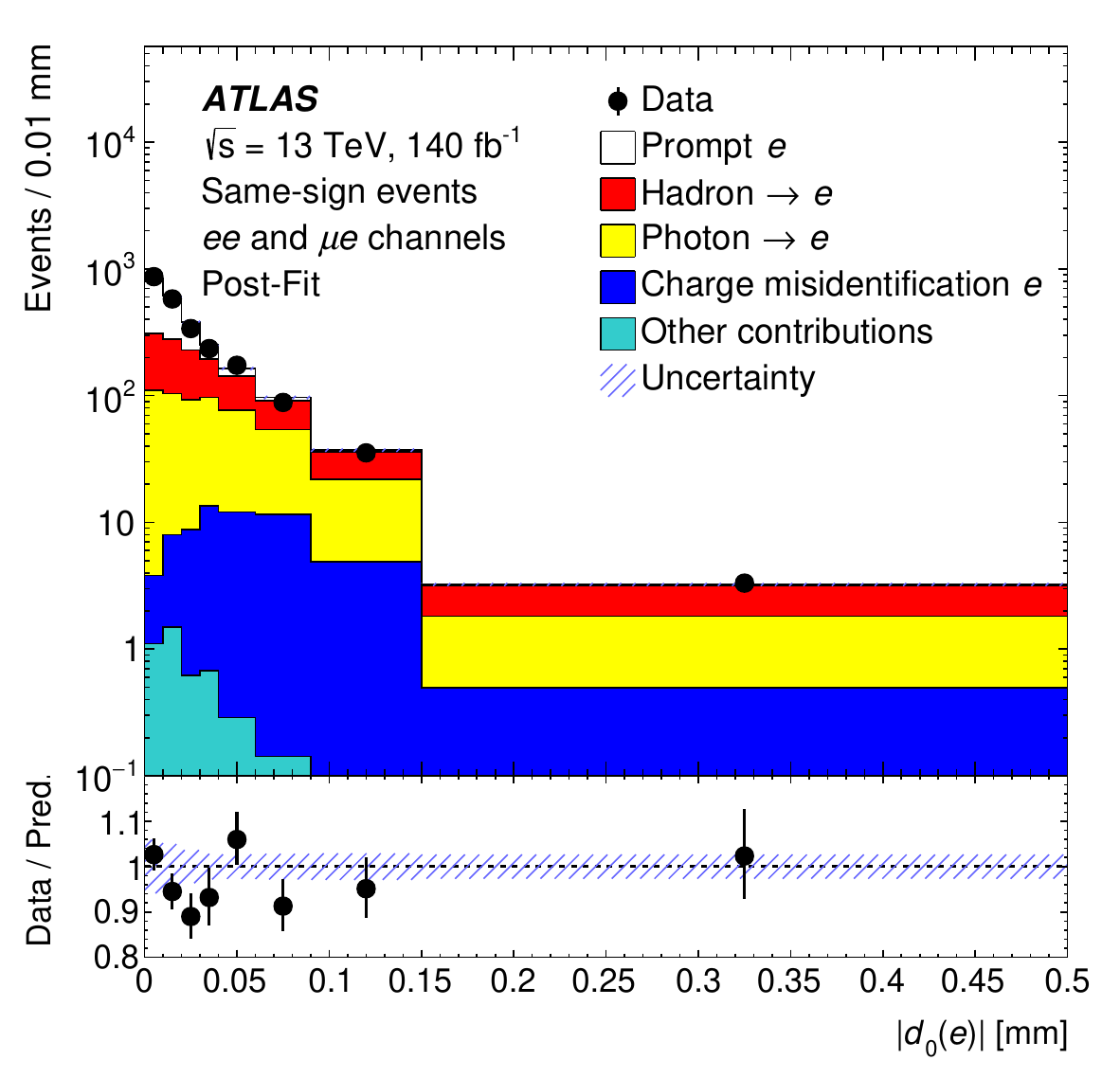}}
\end{center}
\caption{Distributions of (a) \pT  and (b) $|\dzero|$ of the probe electrons for the sum of the $\mu e$ and $e e$ channels of the same-sign sample. The last bin in the \pT distribution includes all events with $20 < \pT(e) < 250$ GeV. The data are represented by black markers and the different components contributing to the sample, taken from simulation after the fit (`Post-Fit'), are shown by stacked histograms. The component `Other contributions' contains events with both fake leptons or with charge misidentification of both leptons. Distributions are shown after the fit to the same-sign data has been performed. The systematic uncertainties related to the limited size of the simulation samples and the cross-section of $\ttbar V$ production are represented by the blue hatched bands. The bottom panels show the ratio of the data to the simulation prediction.
}
\label{fig:SS_d0_pt}
\end{figure}

Figure \ref{fig:SS_d0_pt} shows the \pT and $|\dzero|$ distributions of the probe electrons after the fit to the same-sign data. A reasonably good agreement between data and simulation is observed after applying the measured correction factors. The remaining differences are in the regions where the contribution of prompt electrons from $\ttbar V$ and $VV$ production is dominant. These differences are taken into account by the uncertainties in the $\ttbar V$ and $VV$ production together with the systematic uncertainties of fake-lepton contribution.

\FloatBarrier

%
%

% End of text imported from the .//sections/07_Background.tex input file

% The next lines are included from the .//sections/08_Fit.tex input file
\section{Statistical analysis and systematic uncertainties}
\label{sec:fit}

To measure \Rtaue, a two-dimensional binned template profile likelihood fit is performed to the $\pT$ and $|\dzero|$ distributions. The bin boundaries are optimised to provide the best sensitivity to \Rtaue given the available data. This optimisation resulted in three bins in \pT ($7 < \pT < 10$ GeV, $10 < \pT < 20$ GeV, and $20 < \pT < 250$ GeV) and eight bins in $|\dzero|$ ($|\dzero| < 0.01$ mm, $0.01 < |\dzero| < 0.02$ mm, $0.02 < |\dzero| < 0.03$ mm, $0.03 < |\dzero| < 0.04$ mm, $0.04 < |\dzero| < 0.06$ mm, $0.06 < |\dzero| < 0.09$ mm, $0.09 < |\dzero| < 0.15$ mm, $0.15 < |\dzero| < 0.5$ mm)
in both the $e e$ and $\mu e$ channels. In total 48 bins are used in the fit.

The fit is set up such that a
negative-log-likelihood minimization is performed with the parameter of interest, $\Rtaue$, two other unknown
parameters $k_{\textrm{sig}}$ and $k(\mu/e)$ defined below, and systematic uncertainties included as nuisance parameters (NP), $\bm{\theta}$. The likelihood function is defined as a product of Poisson probabilities for each bin in each channel and a probability density function for systematic uncertainties
\begin{equation}
L(\bm{n},\bm{\theta}^{0}|\Rtaue, k_{\textrm{sig}}, k(\mu/e), \bm{\theta}) = \prod_{i\in\textrm{bins}} P(n_i|\Rtaue,k_{\textrm{sig}},k(\mu/e),\bm{\theta}) \times  \prod_{j\in\textrm{NPs}} G(\theta^{0}_{j}|\theta_{j}).
\end{equation}
Here $n_i$ is the number of events in bin $i$ and $G$ is a
Gaussian distribution that constrains the nuisance parameter $\theta_j$ to the nominal value $\theta^0_j$.

The scaling factors $\Rtaue$, $k_{\textrm{sig}}, k(\mu/e)$ and more than 300 nuisance parameters representing various statistical and systematic uncertainties are varied in the fit.
The parameter of interest \Rtaue is
applied to the \Wtotaue  events.
The SM value of $\Rtaue = 1$ is set in the MC simulation.
The parameter $k_{\textrm{sig}}$ is the normalisation scaling factor of the \ttbar and $Wt$ processes contributing to the signal. It is applied to both the \Wtotaue and \Wtoe components of the signal. The parameter $k(\mu/e)$ takes into account any residual difference in the modelling of the selection efficiencies of the tag muon and electron. It is applied to both the signal and background contributions to the $\mu e$ channel.

The fit is performed after applying the correction factors to the main components of the background events as explained in Section~\ref{sec:background}. Other background processes are normalised to their theoretical cross-sections.

The main systematic uncertainties in \Rtaue are discussed in detail below.

For the measurement of \Rtaue the \ttbar simulation with NNLO re-weighting is used as explained in Section~\ref{sec:samples}. The uncertainty due to re-weighting of the NNLO matrix element calculations is estimated by comparing the result with and without this re-weighting applied.

The uncertainties related to the modelling of \ttbar and $Wt$ signals are estimated by comparing the baseline \POWHEG + \PYTHIA[8] generator to alternative MC generators of \ttbar and $Wt$ production or to modified configurations of the baseline MC generators. The NNLO re-weighting is available only for some of the alternative MC generators. For consistency, the systematic uncertainties related to the \ttbar and $Wt$ modelling are derived from the comparison of the baseline and alternative MC generators without NNLO re-weighting.

To evaluate the uncertainty of \ttbar production due to the \hdamp parameter, the PS and hadronisation model, the NLO matching, and the gluon recoil scheme, the templates obtained with the \POWPY[8] generator are compared with the simulation obtained with the alternative MC samples described in Section~\ref{sec:samples}.

To estimate the sensitivity of \ttbar production to the amount of parton radiation and potential missing higher-order QCD corrections, the renormalisation and factorisation scales of the baseline MC generator are varied up by a factor of 2.0 and down by a factor of 0.5 from the default values.
The uncertainties in initial-state-radiation (ISR) are estimated by using the Var3c eigen-tune variations of the A14 tune~\cite{ATL-PHYS-PUB-2014-021}.
The impact of final-state-radiation (FSR) uncertainties is evaluated using weights which vary the renormalisation scale for QCD emission in the FSR by factors of 0.5 and 2.0, respectively. The difference between up and down variations divided by two is taken as a symmetric uncertainty related to these variations of the MC generators.

The modelling uncertainties of $Wt$ production related to the PS and hadronisation model, \hdamp parameter, NLO matching, ISR and FSR are estimated similarly to \ttbar events.
The uncertainty
related to the interference between the \ttbar and $Wt$ processes is evaluated by comparing the baseline $Wt$ simulation, which was generated using the diagram removal scheme, and the alternative MC sample generated using the diagram subtraction scheme~\cite{Frixione:2008yi,ATL-PHYS-PUB-2016-020} as described in Section~\ref{sec:samples}.

The uncertainty of 5\% in the $Wt$ cross-section is included as a separate contribution to the systematic uncertainty.
The relative uncertainties of the $\ttbar V$ and $VV$ production cross-section are set to 15\% and 30\%, respectively. The relative uncertainties in the $t$- and $s$-channel single-top-quark production are set to 5\%. These uncertainties produce a negligible impact on \Rtaue.

To estimate the systematic uncertainty due to the proton PDFs, 30 nuisance parameters corresponding to
30 different variations of PDF defined
by the PDF4LHC15 set \cite{Butterworth:2015oua} are included in the fit.

The modelling uncertainties are estimated separately for \ttbar and $Wt$ processes, and treated as correlated between these processes and between \Wtoe and \Wtotaue contributions.

To build the \dzero templates of prompt electrons from \Wtoe decays, the $Z$ sample containing mainly $Z \to e^+ e^-$ events is used as explained in Section~\ref{sec:d0_cal}. The electrons from these two processes originate in the hard-scatter primary vertex and therefore their $d_0$ distributions are expected to be the same. However, second-order effects related to track reconstruction and
differences in kinematics of electrons coming from $Z$-boson decay and from \ttbar
result in residual differences between the templates employed in the fit and the actual \dzero distributions. To evaluate the impact of this difference on the result, the \dzero templates are modified as
\begin{equation}
T'_{ij}(\dzero) = N\, T_{i j}^{\textrm{pr}}(\dzero) \frac{D_{ij}^{\Wtoe}(\dzero)}{D_{ij}^{Z}(\dzero)}.
\end{equation}

Here, $D_{ij}^{Z}(\dzero)$ and $D_{ij}^{\Wtoe}(\dzero)$ are the normalised $d_0$ distributions of prompt electrons produced in $Z \to e^+ e^-$ and $t \bar t$ processes, respectively, which are obtained in simulation. The coefficient $N$ normalises the modified templates $T'_{ij}(\dzero)$ to unity. The difference in the value of \Rtaue obtained with the $T'_{ij}(\dzero)$ and $T^{\textrm{pr}}_{ij}(\dzero)$ templates is taken as the systematic uncertainty due to the calibration of $d_0$ of prompt electrons.

Three sources of uncertainty in the evaluation of the correction factors to the number of fake electrons are considered. These include the limited size of the same-sign sample, the choice of PS and hadronisation model, and the possible variation of the fraction of electrons from photon conversion. The uncertainty due to the parton shower and hadronisation model is obtained by comparing the expected number of fake electrons evaluated using the baseline \PYTHIA and alternative \HERWIG algorithms. The fraction of electrons from photon conversions is evaluated using the MC simulation. The effect of the uncertainty in this quantity is estimated by using the fraction in the same-sign sample instead of the nominal opposite-sign sample when measuring \Rtaue.

The electron energy scale and resolution, the reconstruction, isolation, and identification efficiencies and their corresponding uncertainties are measured using $Z \to e^+e^-$ and $J/\psi \to e^+e^-$ events
\cite{ATLAS:2019qmc,ATLAS:2023dxj,EGAM-2021-02-mod}.
In particular, $J/\psi$ events are used to measure the identification efficiency for electrons with $\pT < 20$~GeV. The reconstruction efficiency is not measured for electrons with $\pT < 15$~GeV with the standard tag-and-probe method because of large backgrounds which make the measurement unreliable. The agreement between data and MC in \pT dependence of electron efficiency is essential for this measurement. This agreement is tested using the $Z$ sample of events described in Section~\ref{sec:d0_cal}. The dominant contribution to this sample originates from $Z \to e^+ e^-$ events accompanied with two or more jets.
Following this study, an additional correction to the efficiency of electrons with small \pT
is found to be necessary. The corresponding correction factors are derived from the ratio of the \pT distribution of probe electrons in the $Z$ sample in data and simulation. They are determined separately for 2015--2016, 2017, and 2018 data-taking periods.
The systematic uncertainty due to the additional efficiency correction is evaluated using the comparison of the correction factors obtained with the baseline \SHERPA MC generator and an
alternative sample
generated with \MGNLO.
It is verified that the variation of the correction factors in events with different numbers of jets is within the assigned systematic uncertainty.

Muon reconstruction, identification, and isolation efficiencies, as well as muon momentum scale and resolution, are obtained using the prescriptions specified in Refs. \cite{ATLAS:2020auj,MUON-2022-01}. The size of the corresponding uncertainties are derived using
$J/\psi \to \mu^+ \mu^-$ events for muons with $\pT < 15$ GeV and
$Z \to \mu^+ \mu^-$ events for muons with $\pT > 15$ GeV.

The lepton trigger efficiencies are measured in $Z \to e^+e^-$  and $Z \to \mu^+ \mu^-$ events using
tag-and-probe techniques \cite{ATLAS:2020gty,ATLAS:2019dpa}, and are varied within the corresponding uncertainties.

The jet energy scale and resolution are evaluated using simulation and \textit{in situ} measurements \cite{ATLAS:2020cli}.
For the jet energy scale uncertainty, 35 independent nuisance parameters are used, and 13 nuisance parameters are taken into account for the jet energy resolution.
The uncertainties in the jet $b$-tagging efficiency and rate of $b$-tagged background are estimated by using the semileptonic and dileptonic \ttbar events, as well as $Z+$jets events \cite{ATLAS:2019bwq,ATLAS:2021cxe,ATLAS:2023lwk}.
The simulated \pileup events are weighted to reproduce the
distribution of the average number of inelastic interactions per bunch crossing observed in the data. The associated uncertainty is included as a nuisance parameter in the measurement of \Rtaue.
The uncertainty associated with the use of the JVT algorithm is estimated by using the results of the study presented in Ref. \cite{ATLAS:2015ull}.

The uncertainty in the integrated luminosity is 0.83\% \cite{ATLAS:2022hro}, obtained using the LUCID-2 detector for the primary luminosity measurements, complemented by measurements using the inner detector and calorimeters.
This uncertainty has a negligible impact on \Rtaue.

The measurement of \Rtaue depends on $B(\tau \to e \nu_e \nu_\tau)$, which has been measured to be $(17.82 \pm 0.04)\%$ \cite{Workman:2022ynf}. The dependence on this branching fraction
is included as an additional uncertainty.
The limited size of the MC samples is treated as a separate contribution to the systematic uncertainty following the Barlow-Beeston approach \cite{Barlow:1993dm}.

%

% End of text imported from the .//sections/08_Fit.tex input file

% The next lines are included from the .//sections/09_Results.tex input file
%
%
%
%
%
%
%
%
%
%
%
%
%

\section{Results}
\label{sec:results}

The numbers of observed events in the $\mu e$ and $e e$ channels and in different $\pT$ bins, together with the fitted number of events from different sources, are given in Tables \ref{tab:postyieldmue} and \ref{tab:postyieldee}. The number of events in simulation agrees with that in data across different channels and \pT bins.

\begin{table}[tb]
\begin{center}
\caption{Number of events in the $\mu e$ channel from different sources, as estimated by the fit to the data, compared with the observed yield.
Uncertainties include the statistical and systematic contributions.
The uncertainty in the total expected number of events can be smaller than the uncertainties of the individual contributions because of correlations between them resulting from the fit.}
\label{tab:postyieldmue}
\begin{tabular}{lD{+}{\,\pm\,}{-1}D{+}{\,\pm\,}{-1}D{+}{\,\pm\,}{-1}}
\toprule
& \multicolumn{1}{c}{$\mu e$} & \multicolumn{1}{c}{$\mu e$} & \multicolumn{1}{c}{$\mu e$} \\
& \multicolumn{1}{c}{$7<\pT<10$~GeV} & \multicolumn{1}{c}{$10<\pT<20$~GeV} & \multicolumn{1}{c}{$20<\pT<250$~GeV} \\
\midrule
Prompt $e(\ttbar)$   & 1278\phdo + 28 & 13370\phdo + 150 & 178000\phdo + 1000 \\
$e$ from $\tau$ (\ttbar)   & 1092\phdo + 32 & 4490\phdo + 100 & 11670\phdo + \pho290  \\
Prompt $e(Wt)$   & 34\phdo + \pho6 & 340\phdo + \pho60 & 5300\phdo + \pho900 \\
$e$ from $\tau$ ($Wt$)   & 28.0 + \pho2.5 & 119\phdo + \pho16 & 380\phdo + \pho110  \\
Prompt $e$ (not from \ttbar or $Wt$)   & 5.2 + \pho1.5 & 23\phdo + \phoo7 & 180\phdo + \phoo50 \\
$e$ from $Z \rightarrow \tau^+\tau^-$   & 19.9 + \pho0.4 & 85.4 + \phoo1.4 & 132.9 + \phoo\pho2.2 \\
Fake $e$   & 317\phdo + 22 & 380\phdo + \pho33 & 840\phdo + \phoo60   \\
\midrule
Total predicted  & 2770\phdo + 40 & 18800\phdo + 120 & 196500\phdo + \pho400 \\
\midrule
Data   &  \multicolumn{1}{c}{2768} & \multicolumn{1}{c}{18783} & \multicolumn{1}{c}{196552} \\
\bottomrule
\end{tabular}
\end{center}
\end{table}

\begin{table}[tb]
\begin{center}
\caption{Number of events in the $e e$ channel from different sources, as estimated by the fit to the data, compared with the observed yield.
Uncertainties include the statistical and systematic contributions.
The uncertainty in the total expected number of events can be smaller than the uncertainties of the individual contributions because of correlations between them resulting from the fit.
}
\label{tab:postyieldee}
\begin{tabular}{lD{+}{\,\pm\,}{-1}D{+}{\,\pm\,}{-1}D{+}{\,\pm\,}{-1}}
\toprule
& \multicolumn{1}{c}{$e e$} & \multicolumn{1}{c}{$e e$} & \multicolumn{1}{c}{$e e$} \\
& \multicolumn{1}{c}{$7<\pT<10$~GeV} & \multicolumn{1}{c}{$10<\pT<20$~GeV} & \multicolumn{1}{c}{$20<\pT<250$~GeV} \\
\midrule
Prompt $e(\ttbar)$  & 1238\phdo + 35 & 12210\phdo + 130 & 160300\phdo + 900 \\
$e$ from $\tau$ (\ttbar)   & 1051\phdo + 30 & 4060\phdo + 100 & 10490\phdo + 260 \\
Prompt $e(Wt)$  & 35\phdo + \pho7 & 320\phdo + \pho50 & 5000\phdo + 700 \\
$e$ from $\tau$ ($Wt$)   & 30\phdo + \pho4 & 116\phdo + \pho13 & 340\phdo + 100  \\
$e$ from $Z \rightarrow e^+e^-$   & 240\phdo + 50 & 1770\phdo + 120 & 12380\phdo + 200 \\
Prompt $e$ (not from \ttbar or $Wt$)   & 11.7 + \pho3.5 & 59\phdo + \pho17 & 560\phdo + 170 \\
$e$ from $Z \rightarrow \tau^+\tau^-$   & 19.7 + \pho0.4 & 69.7 + \phoo0.9 & 105.3 + \phoo1.3 \\
Fake $e$   & 302\phdo + 20 & 374\phdo+ \pho32 & 810\phdo + \pho50 \\
\midrule
Total predicted & 2930\phdo + 50 & 18970\phdo + 120 & 190000\phdo + 400 \\
\midrule
Data   & \multicolumn{1}{c}{2928} & \multicolumn{1}{c}{19047} & \multicolumn{1}{c}{189945} \\
\bottomrule
\end{tabular}
\end{center}
\end{table}

\begin{figure}[tbp]
\centering
\subfloat[]{\includegraphics[width=0.49\textwidth]{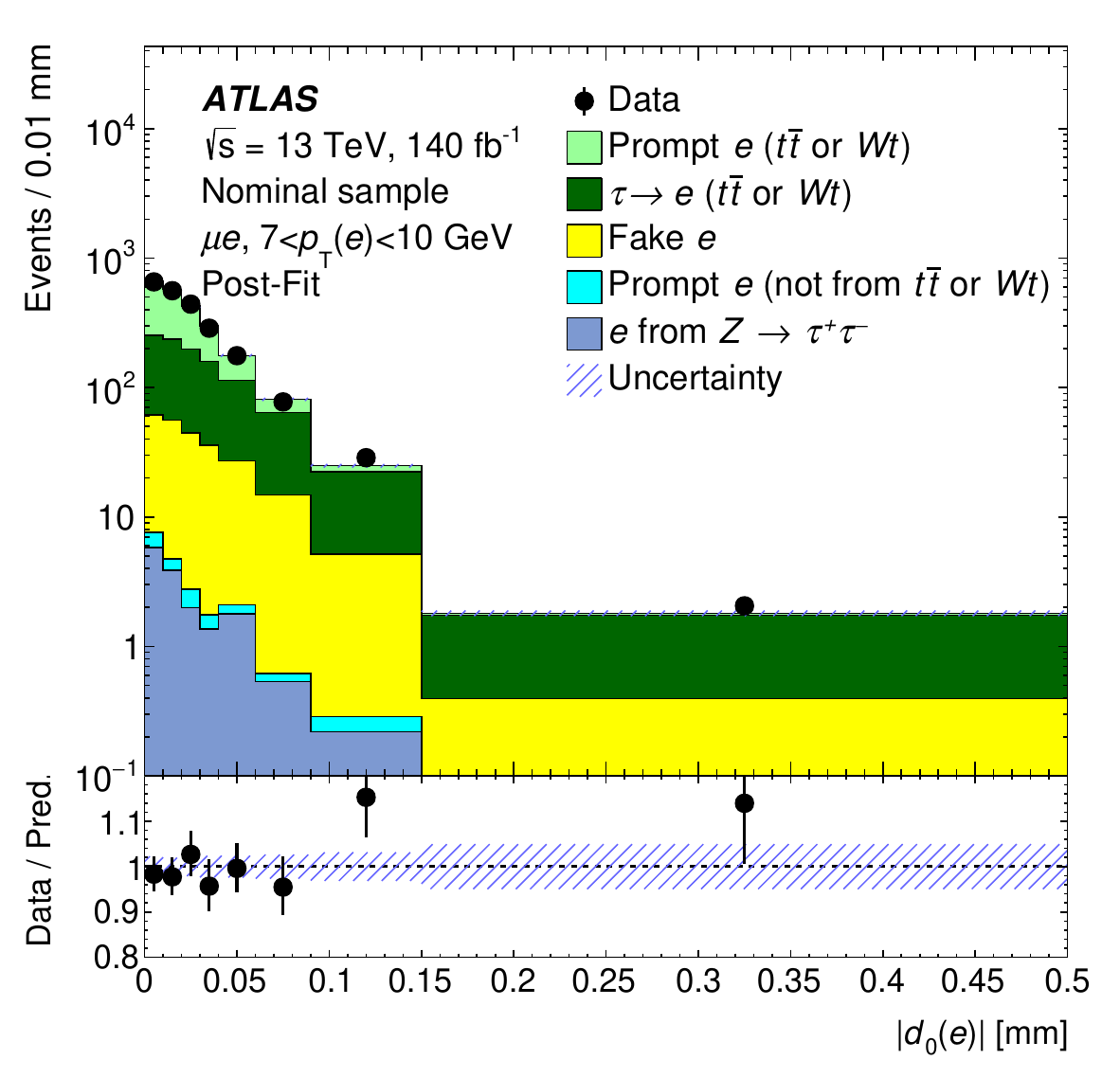}}
\subfloat[]{\includegraphics[width=0.49\textwidth]{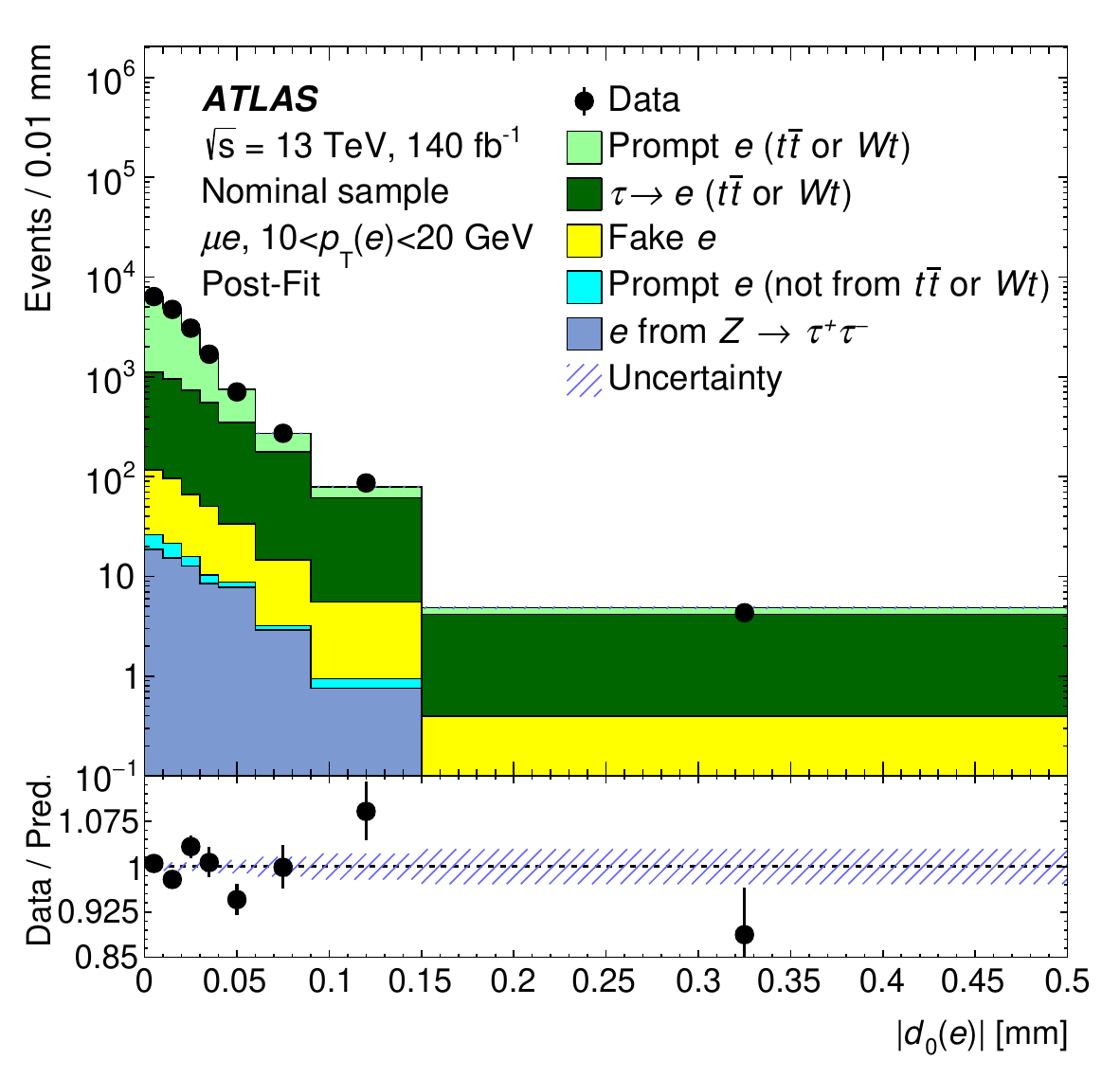}}
\\
\subfloat[]{\includegraphics[width=0.49\textwidth]{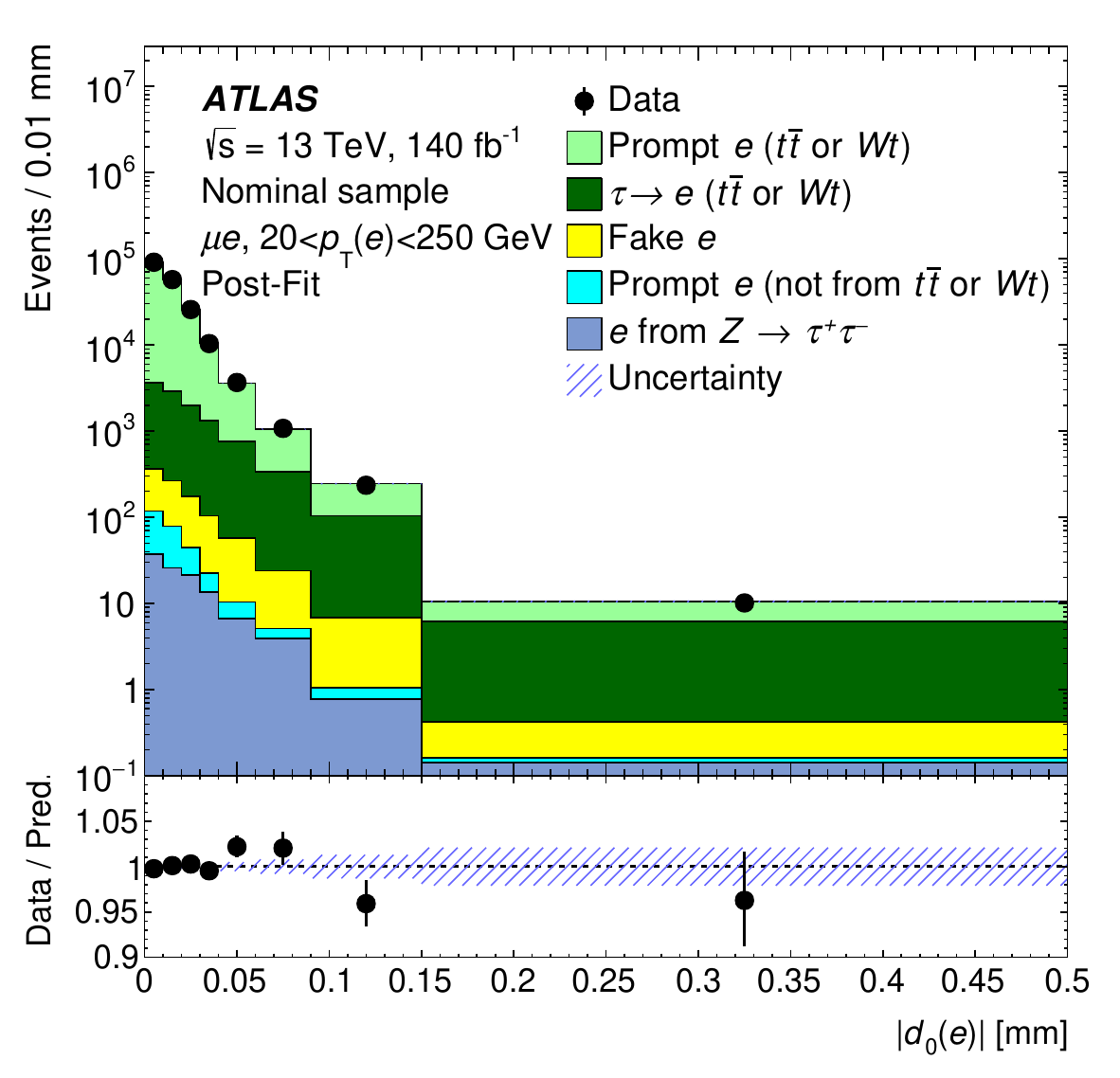}}
\caption{Distributions of $|\dzero|$ for the $\mu e$ channel of the nominal sample in different $\pT$ bins. The data are represented by black markers. The different contributions to the nominal sample, taken from simulation after the fit (`Post-Fit'), are shown by stacked histograms. The bottom panels show the ratio of the data to the simulation prediction after the fit. The systematic uncertainties in the prediction with the constraints from the fit and the correlation between the nuisance parameters applied are represented by the blue hatched bands.
}
\label{fig:postfitd0-me}
\end{figure}

\begin{figure}[tbp]
\centering
\subfloat[]{\includegraphics[width=0.49\textwidth]{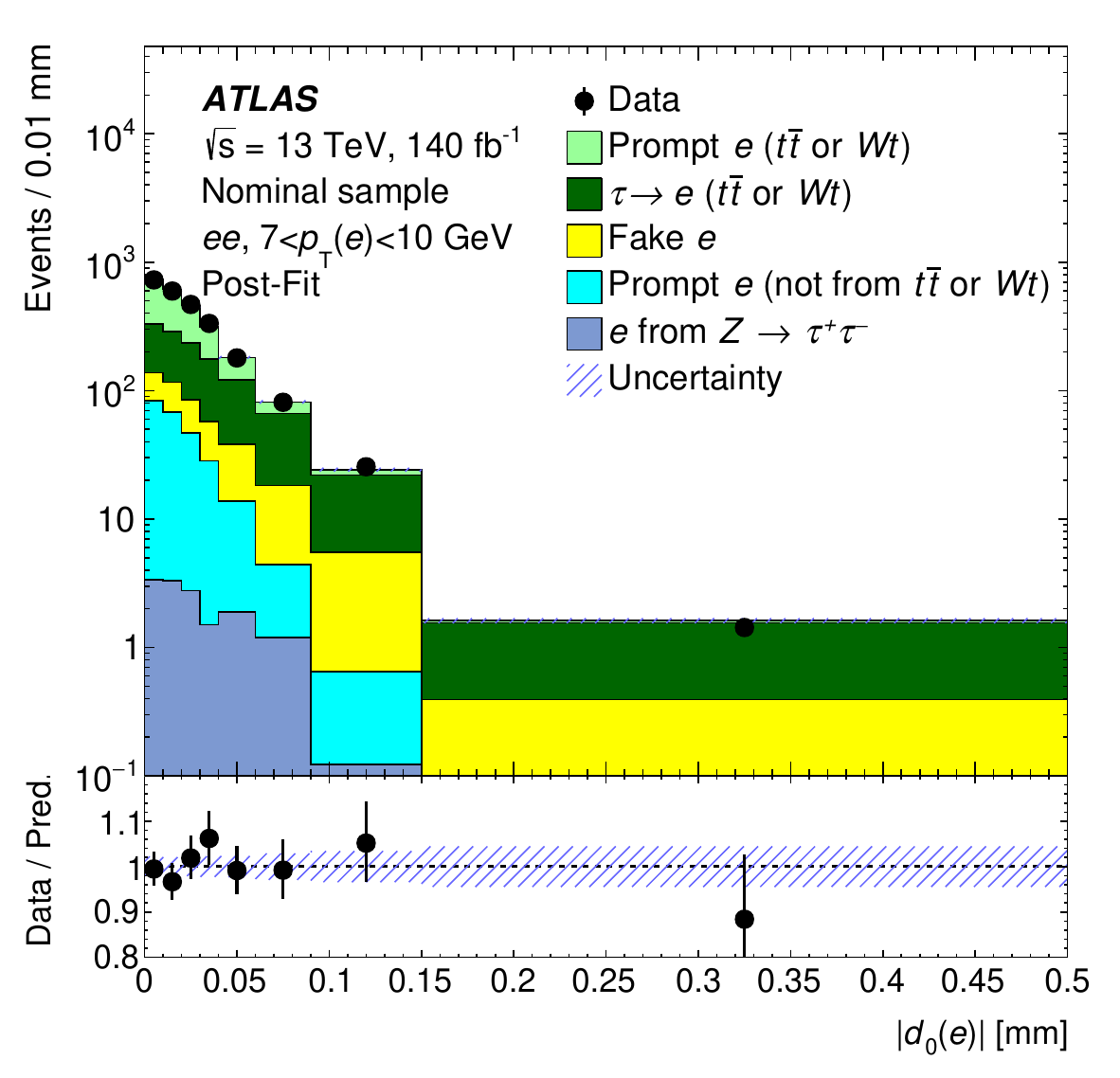}}
\subfloat[]{\includegraphics[width=0.49\textwidth]{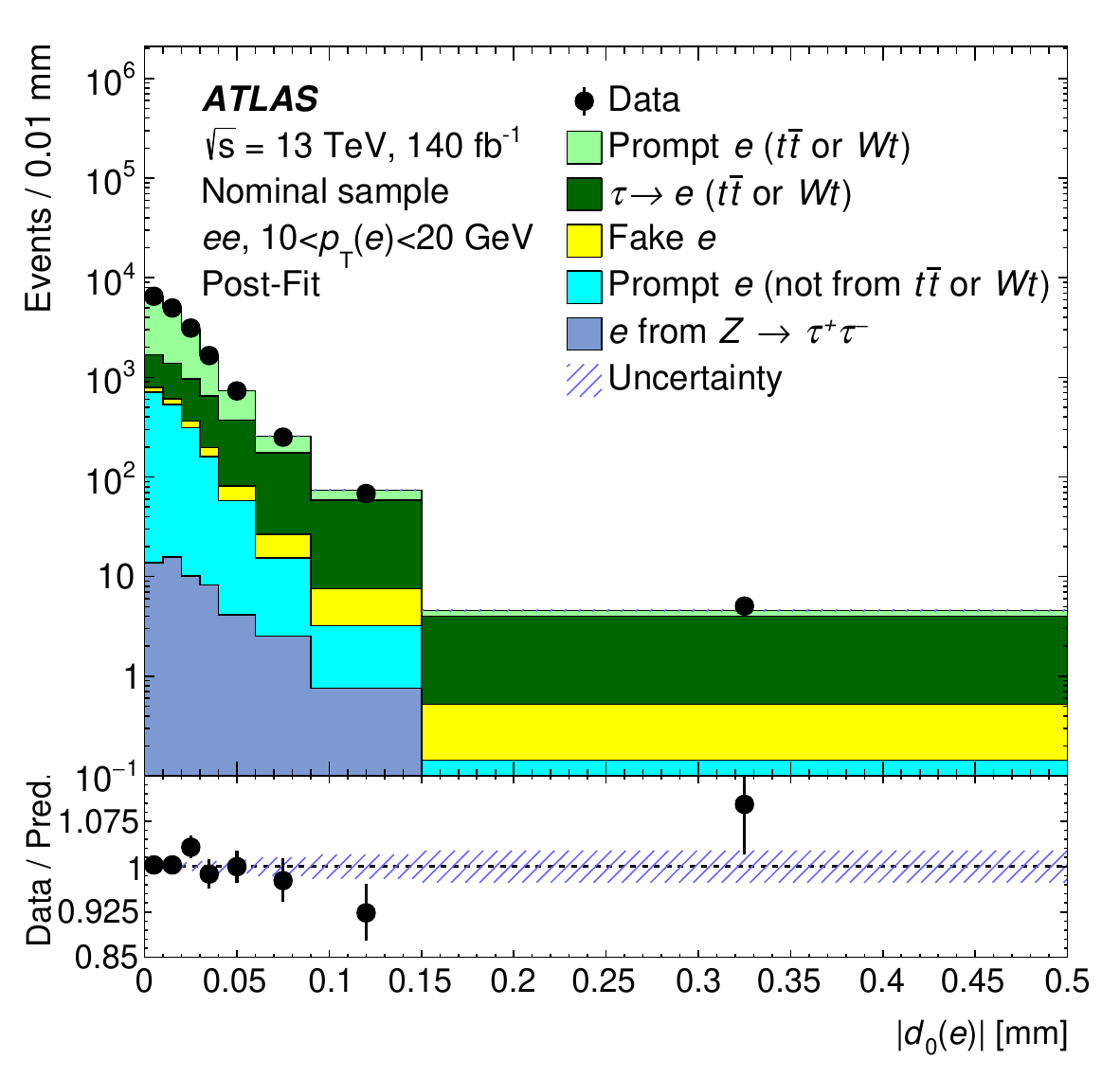}}
\\
\subfloat[]{\includegraphics[width=0.49\textwidth]{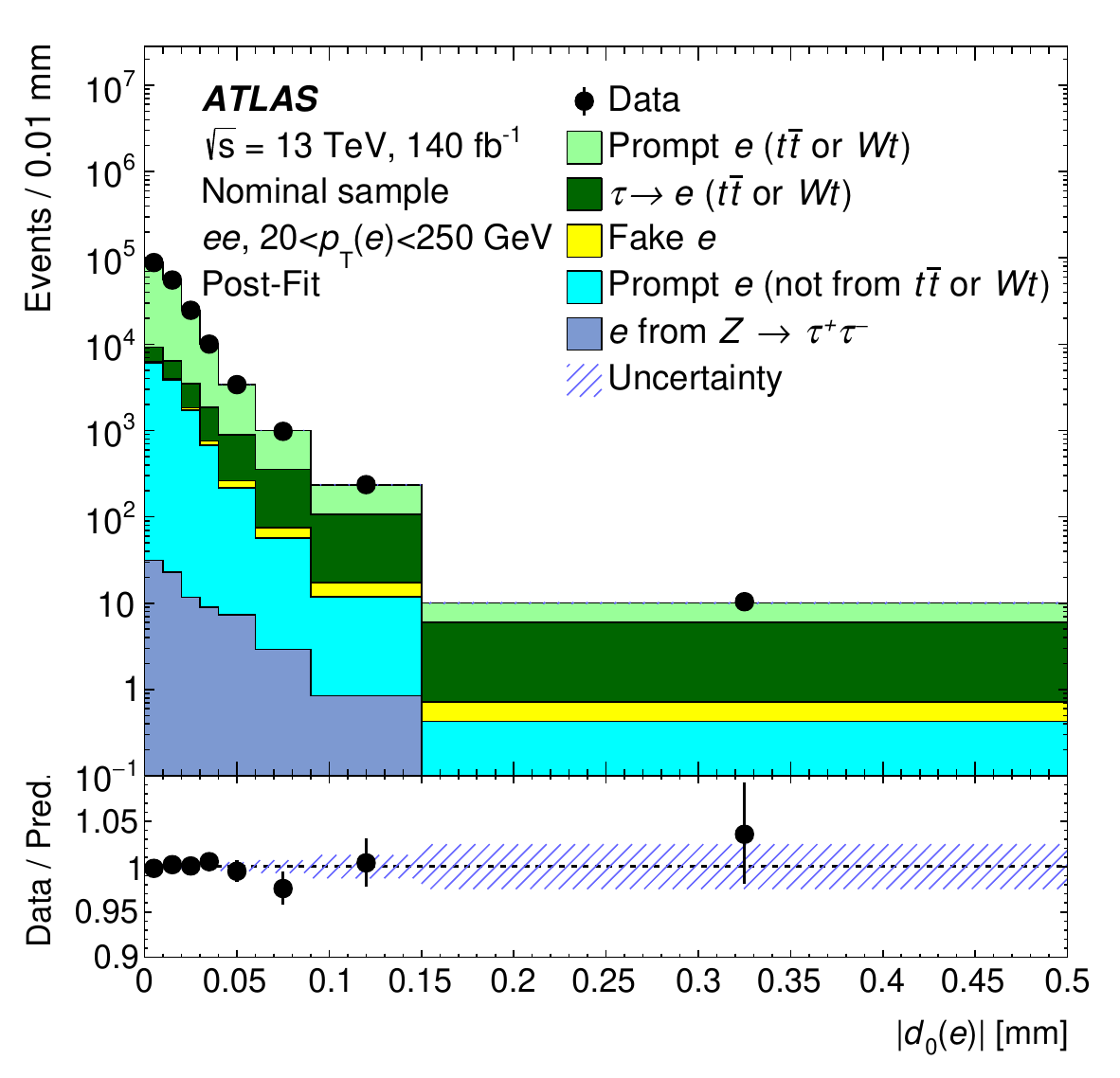}}
\caption{Distributions of $|\dzero|$ for the $e e$ channel of the nominal sample in different $\pT$ bins. The data are represented by black markers. The different contributions to the nominal sample, taken from simulation after the fit (`Post-Fit'), are shown by stacked histograms. The bottom panels show the ratio of the data to the simulation prediction after the fit. The systematic uncertainties in the prediction with the constraints from the fit and the correlation between the nuisance parameters applied are represented by the blue hatched bands.
}
\label{fig:postfitd0-ee}
\end{figure}

Figures \ref{fig:postfitd0-me} and \ref{fig:postfitd0-ee} show the distribution in data and in simulation of $|\dzero|$ in the six regions after the fit.
Good agreement between data and simulation is observed, both in the total yield and differential shape. The $p$-value of the global goodness of the fit \cite{cousins} is 87\%.
These figures
also demonstrate the separation between the signal and background processes. The electrons from \Wtoe dominate at high \pT and small $|\dzero|$, while the electrons from \Wtotaue dominate at high $|\dzero|$. The fake electrons mainly contribute at small \pT.

The measured value of \Rtaue is
\begin{equation}
\Rtaue = 0.975 \pm 0.012~\textrm{(stat.)} \pm 0.020~\textrm{(syst.)}.
\end{equation}

The values of $k_{\textrm{sig}}$ and $k(\mu/e)$ obtained in the fit are $k_{\textrm{sig}} = 1.03 \pm 0.06$ and $k(\mu/e) = 0.986 \pm 0.010$ with uncertainties that include both statistical and systematic components. Both $k_{\textrm{sig}}$ and $k(\mu/e)$ are consistent with unity within less than two standard deviations.
Fixing $k(\mu/e)$ to unity in the fit has a negligible impact on the \Rtaue result, both central value and uncertainties.
Consistent results are found when performing the measurement separately for the 2015--2016, 2017 and 2018 data-taking periods, as well as for the individual channels and $\pT$ bins. Table \ref{tab:XcheckMu:pTbins} shows the results for the different $\pT$ bins.
\begin{table}[tb]
\begin{center}
\caption{Measured values of \Rtaue in the different $\pT$~bins.
}
\label{tab:XcheckMu:pTbins}
\begin{tabular}{l c}
\toprule
\pT bin & $\Rtaue$ \\
\midrule
$7<\pT<10$~GeV & $1.13 \pm 0.11~\textrm{(stat)} \pm 0.07~\textrm{(syst)}$ \\
$10<\pT<20$~GeV & $0.93 \pm  0.04~\textrm{(stat)} \pm 0.02~\textrm{(syst)}$ \\
$20<\pT<250$~GeV & $0.98 \pm  0.04~\textrm{(stat)} \pm 0.02~\textrm{(syst)}$ \\
\bottomrule
\end{tabular}
\end{center}
\end{table}

A breakdown of uncertainties grouped into categories is shown in Table \ref{tab:uncbreak}. Many systematic uncertainties are treated as correlated between the \Wtoe and \Wtotaue contributions. These include the uncertainties in the jet reconstruction, selection of $b$-jets, and trigger efficiencies. The impact of such correlated uncertainties on \Rtaue is reduced.
The remaining dominant uncertainties are related to the modelling of \ttbar production,  the calibration of \dzero distributions (Section \ref{sec:d0_cal}), the evaluation of the background (Section~\ref{sec:background}), and electron reconstruction, identification, and isolation.
The largest impact on \Rtaue in the modelling group of uncertainties comes from NNLO re-weighting, the variation of PS and hadronisation model, the amount of parton radiation, and \hdamp parameter.

\begin{table}[htbp]
\begin{center}
\caption{\label{tab:uncbreak} Breakdown of statistical and systematic uncertainties in the fit to data. Different individual components used in the fit are combined into categories. The size of the impact each group of uncertainties has on \Rtaue, which is denoted by $\sigma(\Rtaue)$ and shown in the second column, is assessed by subtracting in quadrature the uncertainty from the nominal fit and the fit with the relevant nuisance parameters fixed to their post-fit values. %
The group `Other sources' includes the uncertainties due to trigger, JVT, luminosity, and \pileup.
}
\begin{tabular}{lc}
\toprule
Uncertainty group         & $\sigma(\Rtaue)$ \\
\midrule
Modelling of \ttbar and $Wt$  & 0.011 \\
\dzero calibration            & 0.006 \\
Background estimation         & 0.005 \\
Electron reconstruction, identification, and isolation      & 0.005 \\
Electron energy scale        & 0.003 \\
Electron energy resolution   & 0.002 \\
Jet energy resolution  & 0.004 \\
Jet energy scale        & 0.003 \\
Jet $b$-tagging        & 0.002 \\
Muon reconstruction, identification, and isolation  & 0.001 \\
Other sources           & 0.002 \\
Variation of $k_{\textrm{sig}}$ and $k(\mu/e)$   & 0.003 \\
Finite size of simulated samples   & 0.003 \\
$B(W\rightarrow\tau\nu_\tau \rightarrow e \nu_e \nu_\tau \nu_\tau)$ & 0.002 \\
\midrule
Total systematical uncertainty        & 0.020 \\
\midrule
Data statistical uncertainty           & 0.012 \\
\midrule
Total uncertainty         & 0.024 \\
\bottomrule
\end{tabular}
\vspace*{0.1cm}
\end{center}
\end{table}

\begin{figure}[htbp]
\centering
\includegraphics[width=0.90\textwidth]{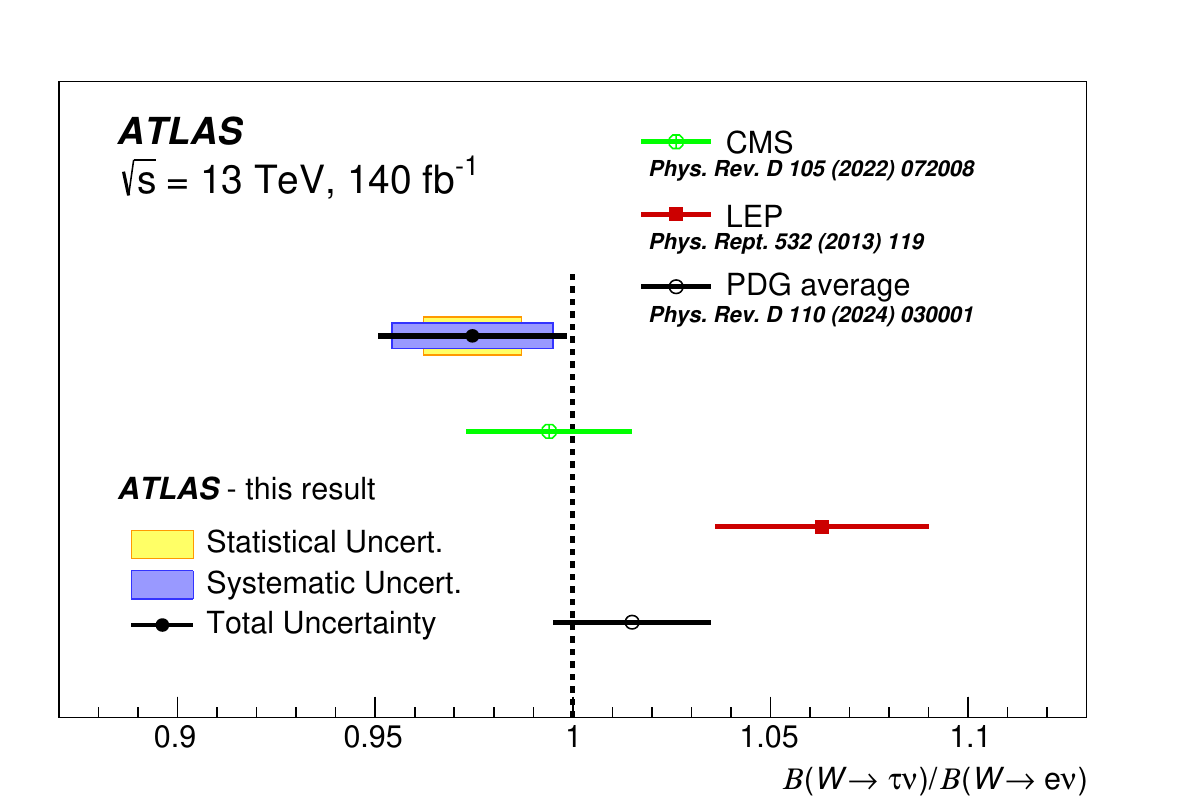}
\caption{Summary of measurements of $\Rtaue = B(W \to \tau \nu_\tau) / B(W \to e \nu_e)$. The \Rtaue value measured in this publication is shown together with the corresponding statistical and systematic uncertainties. It is compared with other measurements and the average value from the PDG \cite{Workman:2022ynf}. The vertical dashed line at unity indicates the Standard Model assumption of equal branching fractions to all lepton flavours.
}
\label{fig:summary}
\end{figure}

Figure \ref{fig:summary} provides a comparison of this result to other measurements of \Rtaue. A similar precision is achieved to that of the CMS result~\cite{CMS:2022mhs}. This new measurement agrees with the SM expectation of lepton flavour universality and with the previous measurement of \Rtaue by the CMS Collaboration. It however differs from the combination of LEP measurements~\cite{ALEPH:2013dgf} by more than two standard deviations.

\FloatBarrier

\section{Conclusion}
\label{sec:conclusion}

The value of $\Rtaue = B(W \to \tau \nu_{\tau})/B(W \to e \nu_e)$ has been measured using the decays of the top quark to final states containing $\tau$-leptons or electrons. A sample of $pp$ collisions at $\sqrt{s}=13$~TeV collected by the ATLAS experiment in Run 2 and corresponding to 140~fb$^{-1}$ was used in the analysis. The leptonic decay $\tau \to e \nu_e \nu_{\tau}$ was used for the $\tau$-lepton selection. The decays $W \to \tau \nu_{\tau} \to e \nu_e \nu_\tau \nu_\tau$ and $W \to e \nu_e$ were distinguished by exploiting differences in the impact parameter and transverse momentum distributions of the electrons.
The dominant systematic uncertainties are related to the modelling of \ttbar and $Wt$ production, the impact parameter calibration, the estimate of the background, and electron reconstruction, identification, and isolation requirements.
The resulting value of $\Rtaue =  0.975 \pm 0.012~\textrm{(stat)} \pm 0.020~\textrm{(syst)}$ is consistent with the Standard Model assumption of lepton flavour universality.
The precision of this measurement is comparable to that of the LEP combination and that of the CMS Collaboration, and will further reduce the uncertainty in the world-average result for \Rtaue.

% End of text imported from the .//sections/09_Results.tex input file

%
%
%
%
%
\section*{Acknowledgements}
%

%

% The next lines are included from the .//acknowledgements/Acknowledgements.tex input file
%
%

%
%

We thank CERN for the very successful operation of the LHC and its injectors, as well as the support staff at
CERN and at our institutions worldwide without whom ATLAS could not be operated efficiently.

The crucial computing support from all WLCG partners is acknowledged gratefully, in particular from CERN, the ATLAS Tier-1 facilities at TRIUMF/SFU (Canada), NDGF (Denmark, Norway, Sweden), CC-IN2P3 (France), KIT/GridKA (Germany), INFN-CNAF (Italy), NL-T1 (Netherlands), PIC (Spain), RAL (UK) and BNL (USA), the Tier-2 facilities worldwide and large non-WLCG resource providers. Major contributors of computing resources are listed in Ref.~\cite{ATL-SOFT-PUB-2023-001}.

We gratefully acknowledge the support of ANPCyT, Argentina; YerPhI, Armenia; ARC, Australia; BMWFW and FWF, Austria; ANAS, Azerbaijan; CNPq and FAPESP, Brazil; NSERC, NRC and CFI, Canada; CERN; ANID, Chile; CAS, MOST and NSFC, China; Minciencias, Colombia; MEYS CR, Czech Republic; DNRF and DNSRC, Denmark; IN2P3-CNRS and CEA-DRF/IRFU, France; SRNSFG, Georgia; BMBF, HGF and MPG, Germany; GSRI, Greece; RGC and Hong Kong SAR, China; ICHEP and Academy of Sciences and Humanities, Israel; INFN, Italy; MEXT and JSPS, Japan; CNRST, Morocco; NWO, Netherlands; RCN, Norway; MNiSW, Poland; FCT, Portugal; MNE/IFA, Romania; MSTDI, Serbia; MSSR, Slovakia; ARIS and MVZI, Slovenia; DSI/NRF, South Africa; MICIU/AEI, Spain; SRC and Wallenberg Foundation, Sweden; SERI, SNSF and Cantons of Bern and Geneva, Switzerland; NSTC, Taipei; TENMAK, T\"urkiye; STFC/UKRI, United Kingdom; DOE and NSF, United States of America.

Individual groups and members have received support from BCKDF, CANARIE, CRC and DRAC, Canada; CERN-CZ, FORTE and PRIMUS, Czech Republic; COST, ERC, ERDF, Horizon 2020, ICSC-NextGenerationEU and Marie Sk{\l}odowska-Curie Actions, European Union; Investissements d'Avenir Labex, Investissements d'Avenir Idex and ANR, France; DFG and AvH Foundation, Germany; Herakleitos, Thales and Aristeia programmes co-financed by EU-ESF and the Greek NSRF, Greece; BSF-NSF and MINERVA, Israel; NCN and NAWA, Poland; La Caixa Banking Foundation, CERCA Programme Generalitat de Catalunya and PROMETEO and GenT Programmes Generalitat Valenciana, Spain; G\"{o}ran Gustafssons Stiftelse, Sweden; The Royal Society and Leverhulme Trust, United Kingdom.

In addition, individual members wish to acknowledge support from Armenia: Yerevan Physics Institute (FAPERJ); CERN: European Organization for Nuclear Research (CERN PJAS); Chile: Agencia Nacional de Investigaci\'on y Desarrollo (FONDECYT 1230812, FONDECYT 1230987, FONDECYT 1240864); China: Chinese Ministry of Science and Technology (MOST-2023YFA1605700, MOST-2023YFA1609300), National Natural Science Foundation of China (NSFC - 12175119, NSFC 12275265, NSFC-12075060); Czech Republic: Czech Science Foundation (GACR - 24-11373S), Ministry of Education Youth and Sports (FORTE CZ.02.01.01/00/22\_008/0004632), PRIMUS Research Programme (PRIMUS/21/SCI/017); EU: H2020 European Research Council (ERC - 101002463); European Union: European Research Council (ERC - 948254, ERC 101089007), European Union, Future Artificial Intelligence Research (FAIR-NextGenerationEU PE00000013), Italian Center for High Performance Computing, Big Data and Quantum Computing (ICSC, NextGenerationEU); France: Agence Nationale de la Recherche (ANR-20-CE31-0013, ANR-21-CE31-0013, ANR-21-CE31-0022, ANR-22-EDIR-0002); Germany: Baden-Württemberg Stiftung (BW Stiftung-Postdoc Eliteprogramme), Deutsche Forschungsgemeinschaft (DFG - 469666862, DFG - CR 312/5-2); Italy: Istituto Nazionale di Fisica Nucleare (ICSC, NextGenerationEU), Ministero dell'Università e della Ricerca (PRIN - 20223N7F8K - PNRR M4.C2.1.1); Japan: Japan Society for the Promotion of Science (JSPS KAKENHI JP22H01227, JSPS KAKENHI JP22H04944, JSPS KAKENHI JP22KK0227, JSPS KAKENHI JP23KK0245); Netherlands: Netherlands Organisation for Scientific Research (NWO Veni 2020 - VI.Veni.202.179); Norway: Research Council of Norway (RCN-314472); Poland: Ministry of Science and Higher Education (IDUB AGH, POB8, D4 no 9722), Polish National Agency for Academic Exchange (PPN/PPO/2020/1/00002/U/00001), Polish National Science Centre (NCN 2021/42/E/ST2/00350, NCN OPUS 2023/51/B/ST2/02507, NCN OPUS nr 2022/47/B/ST2/03059, NCN UMO-2019/34/E/ST2/00393, NCN \& H2020 MSCA 945339, UMO-2020/37/B/ST2/01043, UMO-2021/40/C/ST2/00187, UMO-2022/47/O/ST2/00148, UMO-2023/49/B/ST2/04085, UMO-2023/51/B/ST2/00920); Slovenia: Slovenian Research Agency (ARIS grant J1-3010); Spain: Generalitat Valenciana (Artemisa, FEDER, IDIFEDER/2018/048), Ministry of Science and Innovation (MCIN \& NextGenEU PCI2022-135018-2, MICIN \& FEDER PID2021-125273NB, RYC2019-028510-I, RYC2020-030254-I, RYC2021-031273-I, RYC2022-038164-I); Sweden: Carl Trygger Foundation (Carl Trygger Foundation CTS 22:2312), Swedish Research Council (Swedish Research Council 2023-04654, VR 2018-00482, VR 2022-03845, VR 2022-04683, VR 2023-03403, VR grant 2021-03651), Knut and Alice Wallenberg Foundation (KAW 2018.0458, KAW 2019.0447, KAW 2022.0358); Switzerland: Swiss National Science Foundation (SNSF - PCEFP2\_194658); United Kingdom: Leverhulme Trust (Leverhulme Trust RPG-2020-004), Royal Society (NIF-R1-231091); United States of America: U.S. Department of Energy (ECA DE-AC02-76SF00515), Neubauer Family Foundation.

%
%

% End of text imported from the .//acknowledgements/Acknowledgements.tex input file

%
\clearpage
\printbibliography
\clearpage
\input{atlas_authlist}
%

%
%

%

%
%
%
%

\end{document}

%% file: atlas_authlist.tex
% ATLAS Collaboration author list
% Reference date of TOPQ-2023-10 is 2024-08-27
% Author list last updated on date 16-DEC-24
% Data extracted on 16-Dec-2024 for paper reference TOPQ-2023-10
% at 9:37am
 
\begin{flushleft}
\hypersetup{urlcolor=black}
{\Large The ATLAS Collaboration}

\bigskip

\AtlasOrcid[0000-0002-6665-4934]{G.~Aad}$^\textrm{\scriptsize 105}$,
\AtlasOrcid[0000-0001-7616-1554]{E.~Aakvaag}$^\textrm{\scriptsize 17}$,
\AtlasOrcid[0000-0002-5888-2734]{B.~Abbott}$^\textrm{\scriptsize 124}$,
\AtlasOrcid[0000-0002-0287-5869]{S.~Abdelhameed}$^\textrm{\scriptsize 120a}$,
\AtlasOrcid[0000-0002-1002-1652]{K.~Abeling}$^\textrm{\scriptsize 57}$,
\AtlasOrcid[0000-0001-5763-2760]{N.J.~Abicht}$^\textrm{\scriptsize 51}$,
\AtlasOrcid[0000-0002-8496-9294]{S.H.~Abidi}$^\textrm{\scriptsize 30}$,
\AtlasOrcid[0009-0003-6578-220X]{M.~Aboelela}$^\textrm{\scriptsize 46}$,
\AtlasOrcid[0000-0002-9987-2292]{A.~Aboulhorma}$^\textrm{\scriptsize 36e}$,
\AtlasOrcid[0000-0001-5329-6640]{H.~Abramowicz}$^\textrm{\scriptsize 156}$
% ...
\bigskip
\\

% ...
$^{17}$Department for Physics and Technology, University of Bergen, Bergen; Norway.\\
% ...
$^{30}$Physics Department, Brookhaven National Laboratory, Upton NY; United States of America.\\
% ...
$^{36}$$^{(a)}$Facult\'e des Sciences Ain Chock, Universit\'e Hassan II de Casablanca;$^{(b)}$Facult\'{e} des Sciences, Universit\'{e} Ibn-Tofail, K\'{e}nitra;$^{(c)}$Facult\'e des Sciences Semlalia, Universit\'e Cadi Ayyad, LPHEA-Marrakech;$^{(d)}$LPMR, Facult\'e des Sciences, Universit\'e Mohamed Premier, Oujda;$^{(e)}$Facult\'e des sciences, Universit\'e Mohammed V, Rabat;$^{(f)}$Institute of Applied Physics, Mohammed VI Polytechnic University, Ben Guerir; Morocco.\\
% ...
$^{46}$Physics Department, Southern Methodist University, Dallas TX; United States of America.\\
% ...
$^{51}$Fakult\"{a}t Physik , Technische Universit{\"a}t Dortmund, Dortmund; Germany.\\
% ...
$^{57}$II. Physikalisches Institut, Georg-August-Universit\"{a}t G\"ottingen, G\"ottingen; Germany.\\
% ...
$^{105}$CPPM, Aix-Marseille Universit\'e, CNRS/IN2P3, Marseille; France.\\
% ...
$^{120}$$^{(a)}$New York University Abu Dhabi, Abu Dhabi;$^{(b)}$United Arab Emirates University, Al Ain; United Arab Emirates.\\
% ...
$^{124}$Homer L. Dodge Department of Physics and Astronomy, University of Oklahoma, Norman OK; United States of America.\\
% ...
$^{156}$Raymond and Beverly Sackler School of Physics and Astronomy, Tel Aviv University, Tel Aviv; Israel.\\
% ...
$^{a}$ Also Affiliated with an institute covered by a cooperation agreement with CERN.\\
$^{b}$ Also at An-Najah National University, Nablus; Palestine.\\
$^{c}$ Also at Borough of Manhattan Community College, City University of New York, New York NY; United States of America.\\
$^{d}$ Also at Center for High Energy Physics, Peking University; China.\\
$^{e}$ Also at Center for Interdisciplinary Research and Innovation (CIRI-AUTH), Thessaloniki; Greece.\\
$^{f}$ Also at CERN, Geneva; Switzerland.\\
% ...

\end{flushleft}

% Created with Glance <Atlas.Glance@cern.ch>